\documentclass[alpha-refs]{wiley-article}

\usepackage{adjustbox}
\papertype{Original Article}

\title{Large-eddy Simulations of the Atmospheric Boundary Layer over an Alpine Glacier: Impact of Synoptic Flow Direction and Governing Processes}

\abbrevs{HEF, Hintereisferner; NR, non-stationarity ratio; HEFEX, Hintereisferner experiment; TKE, turbulence kinetic energy; EC, eddy-covariance}

\author[1\authfn{1}]{Brigitta Goger}
\author[1\authfn{1}]{Ivana Stiperski}
\author[1\authfn{1}]{Lindsey Nicholson}
\author[2\authfn{1}]{Tobias Sauter}

\contrib[\authfn{1}]{Equally contributing authors.}

\affil[1]{Department of Atmospheric and Cryospheric Sciences, University of Innsbruck, Innsbruck, Austria}
\affil[2]{Department of Geography, Friedrich-Alexander-Universität Erlangen-Nürnberg, Erlangen, Germany}

\corraddress{Brigitta Goger, Department of Atmospheric and Cryospheric Sciences, University of Innsbruck, Innrain 52, 6020 Innsbruck, Austria}
\corremail{brigitta.goger@uibk.ac.at}

\fundinginfo{BG acknowledges Austrian Science Foundation (FWF) grant number: I~3841-N32; IS acknowledges FWF grant number: T781-N32; TS acknowledges the Deutsche Forschungsgemeinschaft (DFG) grant number: SA~2339/7-1;} 

\runningauthor{Goger et al.}

\begin{document}

\maketitle
\begin{abstract}
The mass balance of mountain glaciers is of interest for several applications (local hydrology or climate projections), and turbulent fluxes can be an important contributor to glacier surface mass balance during strong melting events. The underlying complex terrain leads to spatial heterogeneity and non-stationarity of turbulent fluxes. Due to the contribution of thermally-induced flows and gravity waves, exchange mechanisms are fully three-dimensional, instead of only vertical. Additionally, glaciers have their own distinct microclimate, governed by a down-glacier katabatic wind, which protects the glacier ice and interacts with the surrounding flows on multiple scales. In this study, we perform large-eddy simulations with the WRF model with $\Delta x=48$\,m to gain insight on the boundary-layer processes over an Alpine valley glacier, the Hintereisferner (HEF). We choose two case studies from a measurement campaign (August 2018) with different synoptic wind directions (South-West and North-West). Model evaluation with an array of eddy-covariance stations on the glacier tongue and surroundings reveals that WRF is able to simulate the general glacier boundary-layer structure. Under southwesterly airflow, the down-glacier wind is supported by the South-Western synoptic wind direction, a stable boundary layer is present over the ice surface, and local processes govern the turbulence kinetic energy production. Under northwesterly airflow, a cross-glacier valley flow and a breaking gravity wave lead strong turbulent mixing and to the subsequent erosion of the glacier boundary layer. Stationarity analyses of the sensible heat flux suggest non-stationary behaviour for both case study days, while non-stationarity is highest on the NW day during the gravity-wave event. These results suggest that the synoptic wind direction has, in addition to upstream topography and the atmospheric stability, a strong impact on whether a local glacier boundary layer can form or not, influencing whether a glacier is able to maintain its own microclimate. 

\end{abstract}

\section{Introduction}
Mountain glaciers are an essential part of the global climate system and monitoring their mass balance is required for a large number of applications, such as sea level rise estimates \citep{MarzeionEtAl_2012_Pastfuturesea}, freshwater availability \citep{KaserEtAl_2010_Contributionpotentialglaciers}, or natural hazard warnings \citep{Kaeaeb_2011_NaturalHazardsAssociated}. Glacier mass balance, at any point on a glacier, is a result of energy, mass, and momentum fluxes at the glacier-atmosphere interface \citep{Hock_2005_Glacier}. While the largest part of melt energy stems from the radiation flux, the turbulent fluxes of sensible and latent heat can be equally large over shorter time intervals (e.g., hours), especially during strong ice melting events. Driven by the wind shear as well as thermal stratification and moisture gradients, these turbulent fluxes are important vertical exchange mechanisms coupling the surface and the atmosphere. However, mountain glaciers are not isolated from their environment. The complex interaction with the surrounding topography causes the turbulent fluxes to be spatially variable and this heterogeneity has a relevant impact on glacier mass balance \citep{Sauter.Galos_2016_Effects}.  \newline
In general, the strength of surface fluxes is connected to surface characteristics and the atmospheric boundary layer (ABL) structure, which is spatially heterogeneous in mountainous environments  \citep{Rotach.Zardi_2007_boundary,Lehner.Rotach_2018_Current,Serafin.etal_2018_Exchange}. Mountain boundary layers are influenced by 
dynamically-driven topographic flows such as gravity waves \citep[e. g.,][]{Vosper.etal_2018_Current} and foehn winds \citep{Haid.2020_Foehncoldpoolinteractions}, but also by thermally-induced circulations such as slope flows and valley winds that form due to differential heating/cooling of the surface compared to the overlying atmosphere \citep{Zardi.Whiteman_2013_Diurnal}. Thermally-induced flows interact on multiple scales and are not isloated from each other; e.g. slope flows can be eroded by the larger-scale up-valley wind \citep{Rotach.etal_2008_Boundary}, while the up-valley wind can be superimposed or eroded by the larger-scale synoptic flow, dependent on its strength, the synoptic flow direction, and the local topography \citep[e. g.,][]{Zaengl_2009_impact,Schmidli.etal_2009_External}.
Over glaciers, daytime thermal contrasts result in down-slope katabatic winds (also called glacier winds) that govern the glacier microclimate and protect the glacier ice from warm-air intrusions from aloft \citep{Broeke.M._1997_Structure,Smeets.etal_2000_Turbulence,Oerlemans_2010_microclimate}. The specific vertical structure of katabatic winds with a pronounced jet maximum, as well as large horizontal variability due to the interaction with other complex terrain flows leads to a three-dimensional glacier boundary-layer structure, so that common boundary-layer assumptions, e.g. those used in Monin-Obukhov similarity theory (stationarity, homogeneity, flatness, constant fluxes, and no subsidence), do not hold \citep[e.g.,][]{Rotach.etal_2017_Investigating,FinniganEtAl_2020_BoundaryLayerFlow,Nicholson.Stiperski.2020_Comparisonturbulentstructures}. \newline
Our understanding of the glacier ABL and microclimates stems from a few field campaigns. Thetersonde observations during the PASTEX experiment conducted on the Pasterze glacier in the Eastern Alps \citep{Broeke.M._1997_Structure} showed that the katabatic wind layer can be up to 30\,m deep and its strength depends on interactions with other thermally-induced flows, such as the up-valley wind, or the ambient synoptic flow \citep{Smeets.etal_1998_Turbulence,Broeke_1997_Momentum}. On cloudy, overcast days, katabatic winds are fairly weak, while on fair weather days, a well-developed down-glacier wind can be found over large valley glaciers \citep{Smeets.etal_2000_Turbulence} and even over small perennial ice fields \citep{Mott.etal_2019_Avalanches}. An observational study over the Vatnaj\"{o}kull glacier in Iceland revealed that the local persistent katabatic winds are only disturbed during strong passing storms \citep{Oerlemans.etal_1999_Glacio}. More recent eddy-covariance (EC) observations over glaciers show that synoptic patterns have a strong impact on the turbulence and surface flux structure over Alpine glaciers \citep{Litt.etal_2017_Surface,Nicholson.Stiperski.2020_Comparisonturbulentstructures}. Furthermore, katabatic winds can be enforced or eroded by mesoscale circulations  \citep{ConwayEtAl_2021_IcefieldBreezesMesoscale}, revealing complex flow structures where the underlying atmospheric processes are still not fully understood. The observations from an array of EC towers in the Hintereisferner experiment (HEFEX) over the Hintereisferner glacier (\"{O}tztal Alps, Austria) showed that the glacier tongue is mostly dominated by down-glacier winds stemming from katabatic forcing, but these katabatic winds are frequently disturbed by lateral flow from the North-West \citep{Mott.etal.2020_Spatio-temporal}. These two situations (katabatic and disturbed) exhibited different surface flux and advection patterns, while the processes behind the frequent disturbances remained unknown, because the observations from the EC stations only give limited information about atmospheric processes aloft. \newline 
Simulations with state-of-the art, high-resolution numerical weather prediction (NWP) models offer a tool to overcome this knowledge gap. However, the spatial scale of boundary-layer processes over mountain glaciers require very fine horizontal grid spacings to resolve the relevant processes, dependent on the phenomenon of interest: \cite{Cuxart_2015_When} suggests horizontal grid spacings of less than 10\,m for the correct simulation of stable boundary layers. Idealized large-eddy simulations (LES) revealed that state-of-the-art models are able to simulate the physical properties of katabatic winds over glaciers under idealized settings \citep{Axelsen.Dop_2009_Large,Axelsen.Dop_2009_Largea}, while a semi-idealized modelling approach with real topography by \cite{Sauter.Galos_2016_Effects} showed that surface flux patterns over a mountain glacier are strongly affected by the ambient wind direction and the local topography. These findings were confirmed by \cite{Bonekamp.etal_2020_Using} using direct numerical simulations (DNS) to investigate the impact of small-scale surface inhomogeneities on the surface flux structure over a debris-covered Himalayan glacier. All these aforementioned studies still have a semi-idealized set-up, with a realistic surface, but with an idealized atmosphere (e.g., prescribed wind conditions), while in reality, quickly changing conditions (e.g., shift of the synoptic wind direction, increasing/decreasing cloudiness) might have a relevant impact on the stationarity and heterogeneity of surface fluxes over mountain glaciers. Real-case fine-scale LES process studies (i. e., $\Delta x<100$\,m) over truly complex terrain are still rare and focus on different atmospheric processes such as precipitation patterns \citep{Gerber.etal_2018_Spatial}, real-time weather prediction for sport events \citep{Liu.etal_2020_Simulation}, or foehn-cold air pool interactions \citep{Umek.etal_2021_Large}.\newline
To our knowledge, no real-case LES of boundary-layer processes over a glacier are available to-date. In this study, we perform LES with the Weather Research and Forecasting (WRF) model at $\Delta x=48$\,m of two case study days in August 2018 with two contrary synoptic wind directions: The first day (August 7, 2018), exhibits persistent down-glacier flow under a South-Western synoptic influence, while on the second day (August 17, 2018), the down-glacier wind is strongly disturbed by cross-glacier flows under a North-Western synoptic influence. The simulations aim to show a realistic representation of specific weather situations over a mountain glacier without idealized simplifications or assumptions. The LES set-up allows a validation of model performance with the EC station data, examination of the atmospheric processes leading to the erosion of the down-glacier wind, and an analysis of the spatial patterns and stationarity of the sensible heat flux on the glacier surface. \newline
The paper is organized as follows: In Section~\ref{sec2}, we introduce the area of interest, the observational data, and the model set-up; followed by Section~\ref{sec3}, where we provide a model validation of the general meteorological variables and analyze the wind field; in Section~\ref{sec4} we discuss the temporal and spatial sensible heat flux structure and the horizontal temperature advection patterns. After the discussion of the results (Sec.~\ref{sec5}), we present the conclusions (Sec.~\ref{sec6}).
\section{Data and Methods}\label{sec2}
\subsection{Area of Interest and Observations}
The study area is centered on the Hintereisferner (HEF), a glacier located in in the \"{O}tztal Alps, Austria. HEF (Fig.~\ref{fig1}a) is a approximately 6.3 km long valley glacier, stretching from its highest point, the Wei\ss kugel mountain (3738 m a. s. l.), to a height of 2460\,m, where the glacier tongue's end is located (data from 2018). Tributary glaciers formerly connected to the main tongue include the Stationsferner (small glacier directly located North of StHE in Fig.~\ref{fig1}a) and the Langtaufererjochferner (small glacier located North-West of hefex-3 in Fig.~\ref{fig1}a). HEF has been a subject of long-term mass balance studies since the 1952/53 season and is therefore marked as one of the World Glacier Monitoring Service's reference glaciers \citep{wgms}.  Furthermore, HEF is part of the Rofental catchment, a long-term hydro-meteorological monitoring area \citep{Strasser.etal_2018_Rofental}. \newline 
The mass balance observations are accompanied by operational meteorological stations in the area: The longest dataset stems from the station at Station Hintereis (StHE, 3026\,m a.s.l.), located on the South-West facing slope observing the general meteorological variables and radiation. Data from StHE was used together  with a now-defunct station close to the glacier tongue for a study of the prevailing wind conditions at HEF and surroundings \citep{Obleitner_1994_Climatological}. A permanent Terrerstrial Laser Scanner \citep{VoordendagEtAl_2021_AUTOMATEDPERMANENTLONG} and an EC tower  measuring turbulent statistics are located on the southern ridge (Im Hinteren Eis, IHE, 3245\,m a.s.l.) since 2016. \newline 
In August 2018, the HEFEX campaign was conducted on HEF with a main focus on studying heat exchange mechanisms over the glacier \citep{Mott.etal.2020_Spatio-temporal}. During the campaign, four additional EC stations were employed on the glacier, forming two transects: a cross-transect downwind of the Langtaufererjochferner, and an along-glacier transect along HEF's tongue. The EC stations were equipped with three measurement levels of 2D sonic anemometers (1.7\,m, 1.9\,m, and 2.9\,m) to capture the katabatic down-glacier wind, furthermore, temperature and humidity sensors were installed at 1.9\,m. The observed turbulence data were rotated using double rotation, detrended, and averaged over 15 minutes \citep[for more details see][]{Aubinet.etal_2012_Eddy,Stiperski.Rotach_2016_Measurement,Mott.etal.2020_Spatio-temporal}. The HEFEX campaign provides a dataset of turbulence data over the glacier for around a month, and an analysis of the time series shows that HEF's glacier tongue is mostly dominated by down-glacier winds. However, during certain situations, the down-glacier wind is disturbed by flow from the North-West, together with warm-air advection. The reason behind these disturbances is still unknown after the detailed observational analysis by \cite{Mott.etal.2020_Spatio-temporal}, therefore we focus our simulations on two specific case study days:  August 7, 2018, where the large-scale synoptic flow is southwesterly and down-glacier flows dominate (SW day hereafter), and August 17, 2018, when a northwesterly synoptic flow is present and the wind direction over the glacier tongue was shifting (NW day hereafter).  
\subsection{Numerical Model}
\begin{figure}[bt]
 \centering
 \includegraphics[width=\textwidth,keepaspectratio]{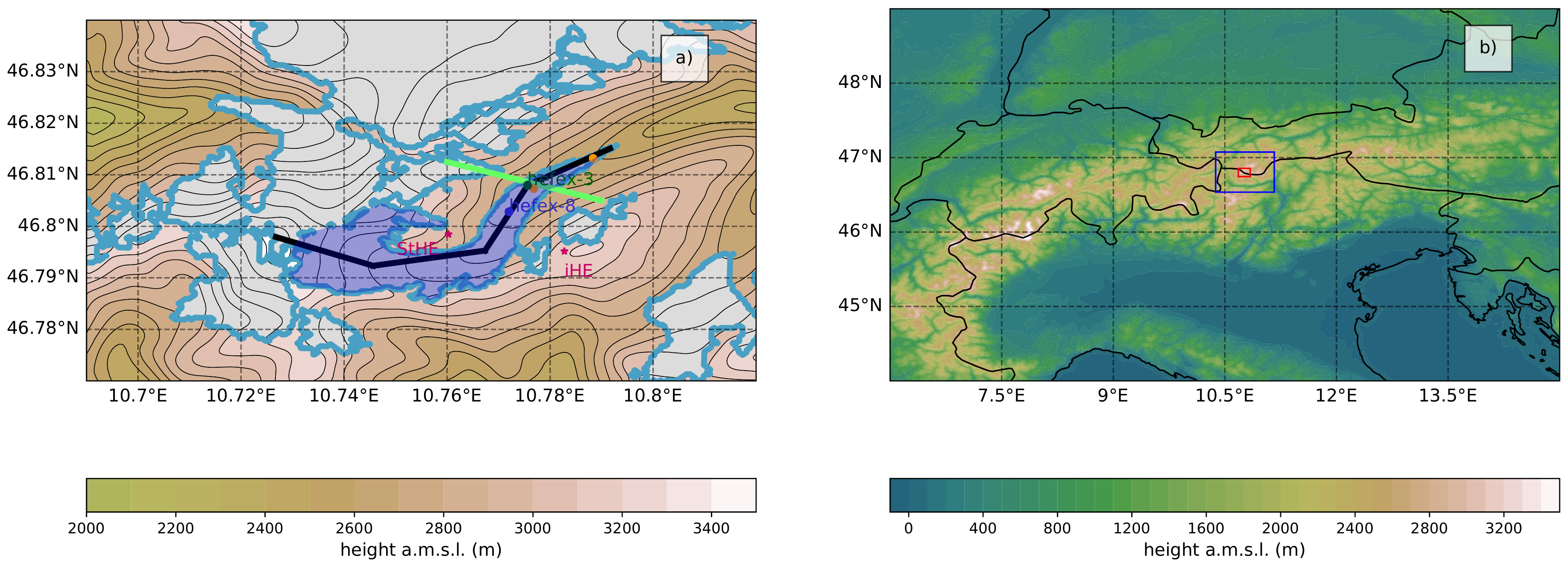}
  \caption{(a) Overview of the innermost model domain: colors and contour lines show the model topography at $\Delta x=48$\,m. The glacier outlines are shown in light blue, while the area of HEF is highlighted in dark blue. Colored points show the permanent (stars) and HEFEX stations. The light-green line shows the cross-section of Fig.~4, while the black line indicates the cross-section of Fig.~5. (b) The mesoscale domain ($\Delta x=1$\,km, where the locations of domain 3 ($\Delta x=240$\,m, blue rectangle) and domain 4 ($\Delta x=48$\,m, red rectangle) are highlighted.}
 \label{fig1}
 \end{figure}

We employ a nested set-up of the Weather Research and Forecasting (WRF) model, version 4.1 \citep{Skamarock.etal_2019_Description} for the two case study days. The ERA5 reanalyses \citep[][$\Delta x\approx 30$\,km]{CCCS_2017_ERA5}, are used as model boundary data every three hours. The set-up consists of four one-way nested domains with $\Delta x=6$\,km (d01, spanning Europe), $\Delta x=1.2$\,km (d02, spanning the Alps, Fig.~\ref{fig1}b), $\Delta x=240$\,m (d03, spanning the \"{O}tztal Alps, red rectangle in Fig.~\ref{fig1}b), and $\Delta x=48$\,m (d04, spanning the Rofental catchment with HEF in the center, Fig.~\ref{fig1}a and blue rectangle in Fig.~\ref{fig1}b), respectively. All domains have 86 vertical levels in terrain-following coordinates, where the lowest model level is located at $z=7$\,m. The coarser-resolution domains d01 and d02 mainly serve as boundary data for the two innermost LES domains. All simulations are initiated at 03 UTC on the respective case study day and are run for 18 hours. \newline
Numerical simulations over complex terrain still face numerous challenges related to static data \citep[e.g., land-use, soil representation, and topography,][]{Meij.Vinuesa_2014_Impact,Goger.etal_2016_Current,Jimenez-Esteve.etal_2018_Land,GolzioEtAl_2021_LandUseImprovements}, therefore  we replaced the model's default static data with higher-resolution datasets: We use the Harmonized World Soil Database \citep{soildatabase} for soil properties, and ESA-CCI land-use data \citep{ESA_2017_LandCoverCCI} with a horizontal grid spacing of 1\,km for land-use categories. The Shuttle Radar Topography Mission (SRTM) 1 Arc-Second Global topography dataset \citep{usgs} is chosen for the model topography, while three cycles of terrain smoothing with the 1-2-1 smoothing filter \citep{GuoChen_1994_Terrainlanduse} are applied. For the two innermost domains with sub-hectometre resolution, it is necessary to replace slopes steeper than 30$^{\circ}$ with slopes from the respective coarser-grid domain to avoid numerical instabilities. This allows the model to keep a part of terrain complexity by reducing the number of terrain smoothing cycles. \newline
 In all domains, the Thompson microphysics scheme \citep{ThompsonEtAl_2008_ExplicitForecastsWinter}, the MM5 revised surface layer scheme \citep{JimenezEtAl_2012_RevisedSchemeWRF}, and the NOAH-MP land-surface scheme \citep{YangEtAl_2011_communityNoahland} are employed. For both long- and shortwave radiation,  the RRTMG scheme  \citep{IaconoEtAl_2008_Radiativeforcinglong} is used while topographic shading is activated. The ABL physics in d01 and d02 are parameterized by the level three Mellor-Yamada-Nakashini-Niino parameterization \citep{Nakanishi.Niino_2009_Development}, because parameterizations of the Mellor-Yamada kind show a good performance for the simulation of thermally-induced circulations \citep{Wagner.etal_2014_Impact,Goger.etal_2018_Impact}. For horizontal diffusion a horizontal Smagorinsky first order scheme is employed \citep{Smagorinsky_1963_General}. \newline
After the mesoscale run (d01 and d02) is finished, we use the \texttt{ndown} routine to create the 30 minute boundary data for the standalone LES (d03 and d04). Special adjustments in the static data are made for the LES runs: We use the CORINE dataset \citep{EEA_2017_CopernicusLandService} for the two innermost domains, reclassifying the land-use categories to USGS categories following the procedure of \cite{Pineda.etal_2004_Using}. However, the CORINE dataset with $\Delta x=100$\,m still does not represent HEF and the surrounding glaciers in a satisfactory way. Therefore, we changed the land use index of the \texttt{wrfinput} files for d03 and d04 with up-to date glacier outlines from the Randolph Glacier Inventory \citep{Pfeffer.etal_2014_Randolph}, and for HEF itself, we use a shapefile acquired with data from an airborne laser scanning campaign in 2018 (Rainer Prinz, personal communication). \newline
In the LES set-up, no boundary-layer parameterization is employed because we expect that a relevant part of the turbulence is already resolved. We assume that turbulence is generated at the inflow boundaries of d04 due to the heterogeneous terrain and because d03 is also run in LES mode. However, we regard the first three simulation hours as model-spin-up time and exclude them from data interpretation. We use for both LES domains a 3D subgrid-scale turbulence parameterization after \cite{Deardorff_1980_Stratocumulus} including a prognostic equation for the subgrid-scale TKE. The length scale for the calculation of the eddy viscosity follows the approach by \cite{Schmidli_2013_Daytime}. Instantaneous model output for the standard WRF variables is generated every 15 minutes, while 1-minute instantaneous output is available for selected variables such as wind, temperature, and the surface sensible heat flux. Additionally, we use the online time averaging tool \textit{WRF LES diagnostics} \citep{lukasumek_2020_3901119} to create 15-minute averaged values of various variables for better comparability with the observations. Unless stated otherwise, all model output presented in the next sections are 15 minute averages from the innermost LES domain (d04, $\Delta x=48$\,m). The closest model grid point from the respective stations is determined via Euclidian distance. 
\section{Temperature and Wind Patterns}\label{sec3}
\subsection{Time Evolution of Meteorological Variables}
\begin{figure}[bt]
\centering
\includegraphics[width=12cm]{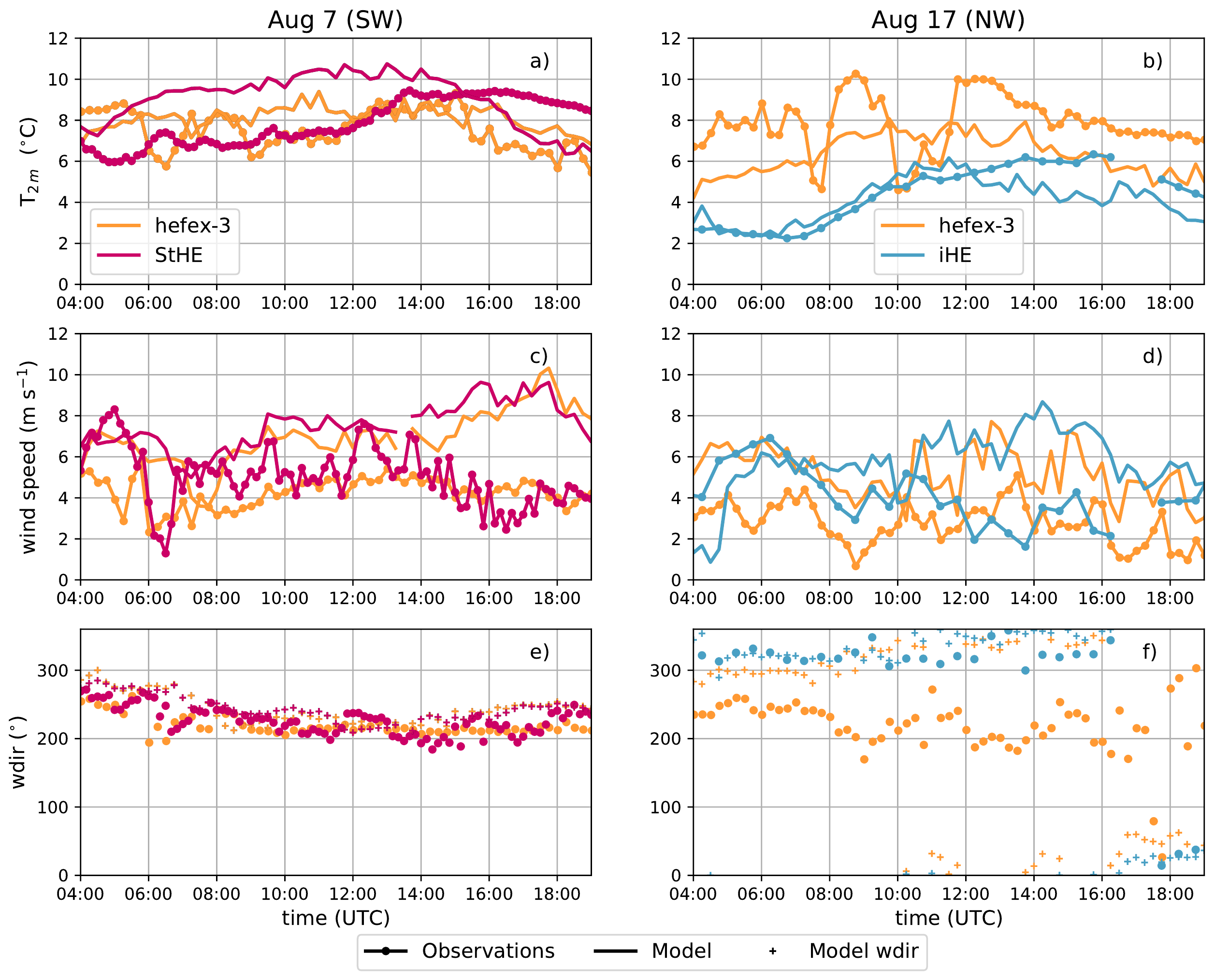}
\caption{Time series of 2\,m temperature (a,b), horizontal wind speed (c,d), and wind direction (e,f) from August 7 (left column) and August 17 (right column). Connected dots indicate observations, while straight lines (or plus signs, respectively) indicate model output. Orange colors show data from the station hefex-3, pink colors indicate StHE, and blue colors indicate IHE.}\label{fig2}
\end{figure}

At first, we perform a classical model evaluation of standard meteorological quantities such as 2\,m temperature, as well as wind speed and direction from the lowest model half-level at 7\,m a. g. to determine whether the model delivers reliable results. The 2\,m temperature over the glacier tongue remains quite constant (around 8$^{\circ}$C) on the SW day (Fig.~\ref{fig2}a). Minor changes are visible for the south-facing slope station StHE, and the model is able to simulate the 2\,m temperature with a slight warm bias (around 2\,$^{\circ}$C). The observed horizontal wind speed (Fig.~\ref{fig2}c) shows differences of less than  1\,m\,s$^{-1}$ between the glacier tongue (hefex-3 station) and the slope station StHE aloft. However, the horizontal wind speed is overestimated by the model for both locations by about 2\,m\,s$^{-1}$ before noon, while the wind speed increases in the afternoon up to 8\,m\,s$^{-1}$ with progressing simulation time. The wind direction at both locations suggests persistent down-glacier flow during the entire day in both the observed and modelled time series (Fig.~\ref{fig2}e). This down-glacier wind direction corresponds to both the larger-scale synoptic flow on this day (South-Westerly), as well as to a possible katabatic glacier wind. However, if the circulation were purely thermally-induced, the wind direction of StHE would show an up-slope wind direction after sunrise (around 180$^{\circ}$) indicating up-slope flows. Since this is is not the case, we conclude that the synoptic flow erodes the small-scale slope flows, but might support the katabatic down-glacier wind. 
\newline 
The second case study day is dominated by North-Westerly synoptic flow, therefore we show, besides the glacier tongue station, timeseries from the ridge EC tower IHE, since its location is representative of the flow across the glacier valley. The temperature differences between these two stations are higher than between hefex-3 and StHE due to the larger height difference (Fig.~\ref{fig2}b). Asides from this, the model is able to simulate the 2\,m temperature fairly well, although after 12\,UTC, there are rapid changes above the glacier surface not captured by the model. The wind speed shows no specific diurnal cycle at both locations, but the model overestimates the wind speed at both stations as on the SW day (Fig.~\ref{fig2}d). However, the wind speed at hefex-3 is less constant with time and quickly changing, especially between 10\;UTC and 12\,UTC. This is also visible in the timeseries of the wind direction (Fig.~\ref{fig2}f): At the glacier tongue, both in the model and in the observations rapid changes in wind direction are present, furthermore, the model's wind direction has an offset of about 50 degrees during the day compared to the observations. According to the model, most of the day, a cross-valley flow is present, while the observations from hefex-3 suggest a heavily disturbed down-glacier flow with quickly-changing wind directions. One reason for this might be the different heights from the observations (around 2\,m) and the height from the lowest model half-level (around 7\,m), suggesting that the model is not able to simulate small-scale processes on the glacier tongue. However, the model simulates the dominating North-Westerly flow at the ridge top station well, asides from the wind speed overestimation. This suggests that the model likely simulates a process related to the cross-glacier flow too strongly over the glacier tongue. None of the wind directions correspond to possible thermally-induced circulations, and we conclude that the synoptic flow dominates the wind patterns over the glacier surface. Since the conditions on this day are strongly dynamically-driven, it is possible that channeling effects or gravity waves are present over the glacier valley, eroding the glacier boundary layer and dominating the small-scale processes. \newline
To summarize, the model is able to simulate the general meteorological variables on the glacier and its surroundings in a realistic way, although we have to point out that the wind speed is systematically overestimated and the fast, chaotic changes in wind direction on the NW day have a directional shift in the model. We decide to focus for both days on the time period from 06\,UTC to 12\,UTC, since during this time frame the model performs the best and this allows a process-based interpretation of model data. If not stated otherwise, all results presented in the following sections are from this timeframe of 6 hours.
\subsection{Horizontal Spatial Patterns of Potential Temperature}\label{sec_pottemp}
\begin{figure}[bt]
\centering
\includegraphics[width=\textwidth,keepaspectratio]{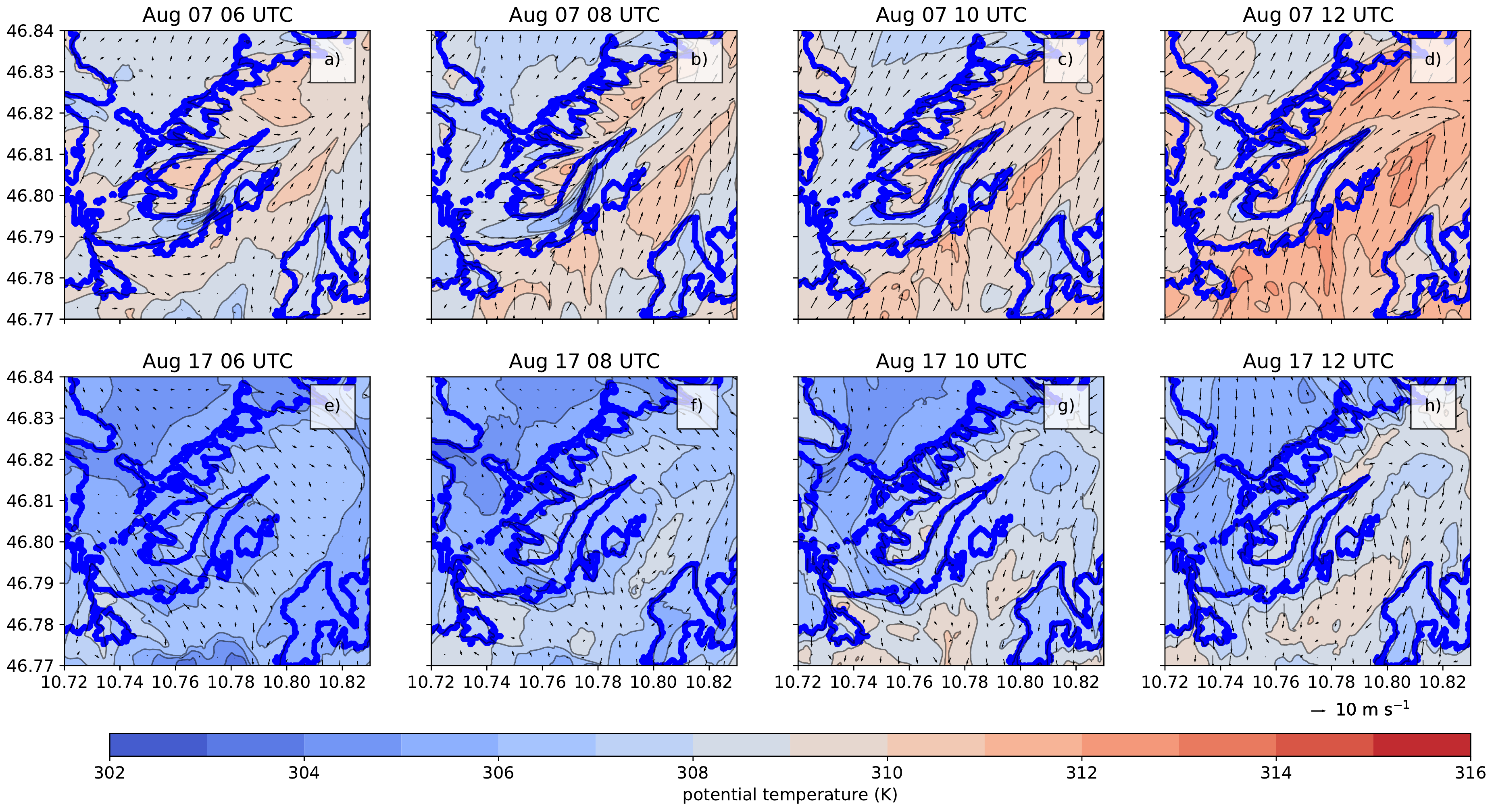}
\caption{Averaged model output from the lowest model level of potential temperature (colors and black contour lines) and horizontal wind vectors. The top row from August 7, while the bottom row shows data from August 17, at four different times. Blue lines indicate the outlines of the glacier as it is represented in the model domain.}\label{fig3}
\end{figure}

The potential temperature provides information about the thermal stratification and, together with the wind field, the spatial variability of the glacier boundary layer structure (Fig.~\ref{fig3}). On early morning of the SW day, the southeast facing slopes are already heated, while potentially colder air is present over the ice surfaces (Fig.~\ref{fig3}a). Plumes of potentially warmer air are transported towards the glacier tongue with the westerly wind, while small plumes of potentially colder air are visible downwind of the smaller tributary glaciers. During the next four hours (Fig.~\ref{fig3}b,c), the synoptic wind direction shifts to South-West, supporting the persistent down-glacier wind. However, when the larger-scale wind direction turns towards southerlies, the body of potentially colder air over the glacier becomes less pronounced (Fig.~\ref{fig3}d) and weakens in the late afternoon (not shown). In general, the SW case study shows a heterogeneous potential temperature structure and large horizontal gradients between the ice surfaces and their surroundings.  \newline
The potential temperature structure is entirely different on the NW day. Although the glacier surfaces are still potentially colder than their surroundings, the horizontal temperature gradients are weaker (Fig.~\ref{fig3}e,f,g,h) and the potential temperature field is spatially homogeneous. During most of the time, a strong North-westerly cross-glacier flow is present, perpendicular to the down-glacier wind direction, thus preventing a small-scale glacier wind to form. The cross-glacier flow often shifts its wind direction slightly, and turns towards up-glacier in the afternoon (Fig.~\ref{fig3}h). The reason for the fast-changing wind direction over the glacier tongue is discussed in the following (Section~\ref{vertstructure}). 
\subsection{Vertical Structure}\label{vertstructure}
\begin{figure}[bt]
\centering
\includegraphics[width=\textwidth,keepaspectratio]{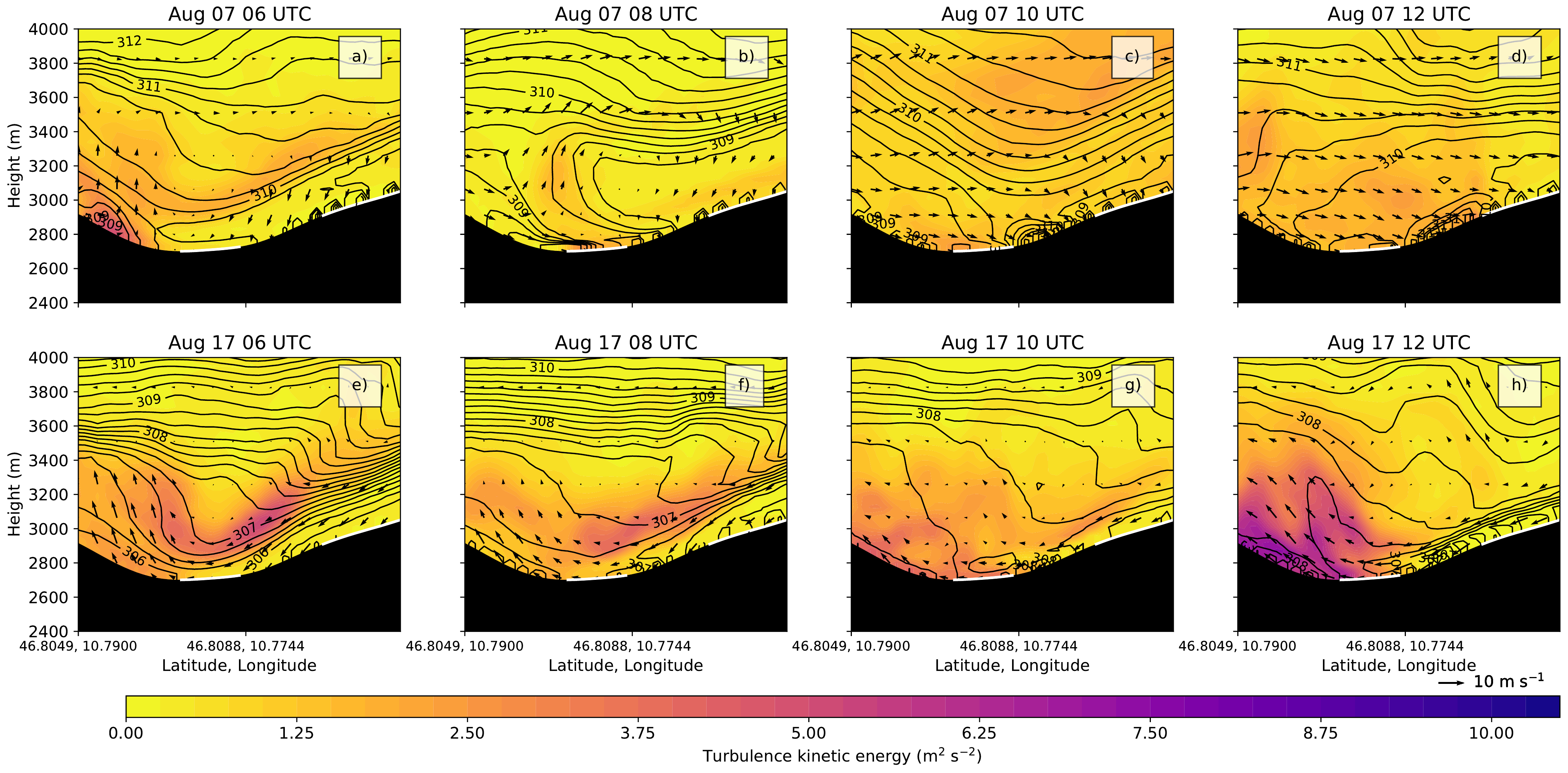}
 \caption{Vertical cross-section along the light-green line in Fig.~1 for the SW day (upper row) and the NW day (lower row). White lines along the topography indicate glaciated areas. Black contour lines (every 0.25\,K) show the potential temperature, wind vectors indicate the cross-valley wind speed, and colors indicate turbulence kinetic energy.}
\label{fig4}
\end{figure}

\begin{figure}[bt]
\centering
\includegraphics[width=\textwidth,keepaspectratio]{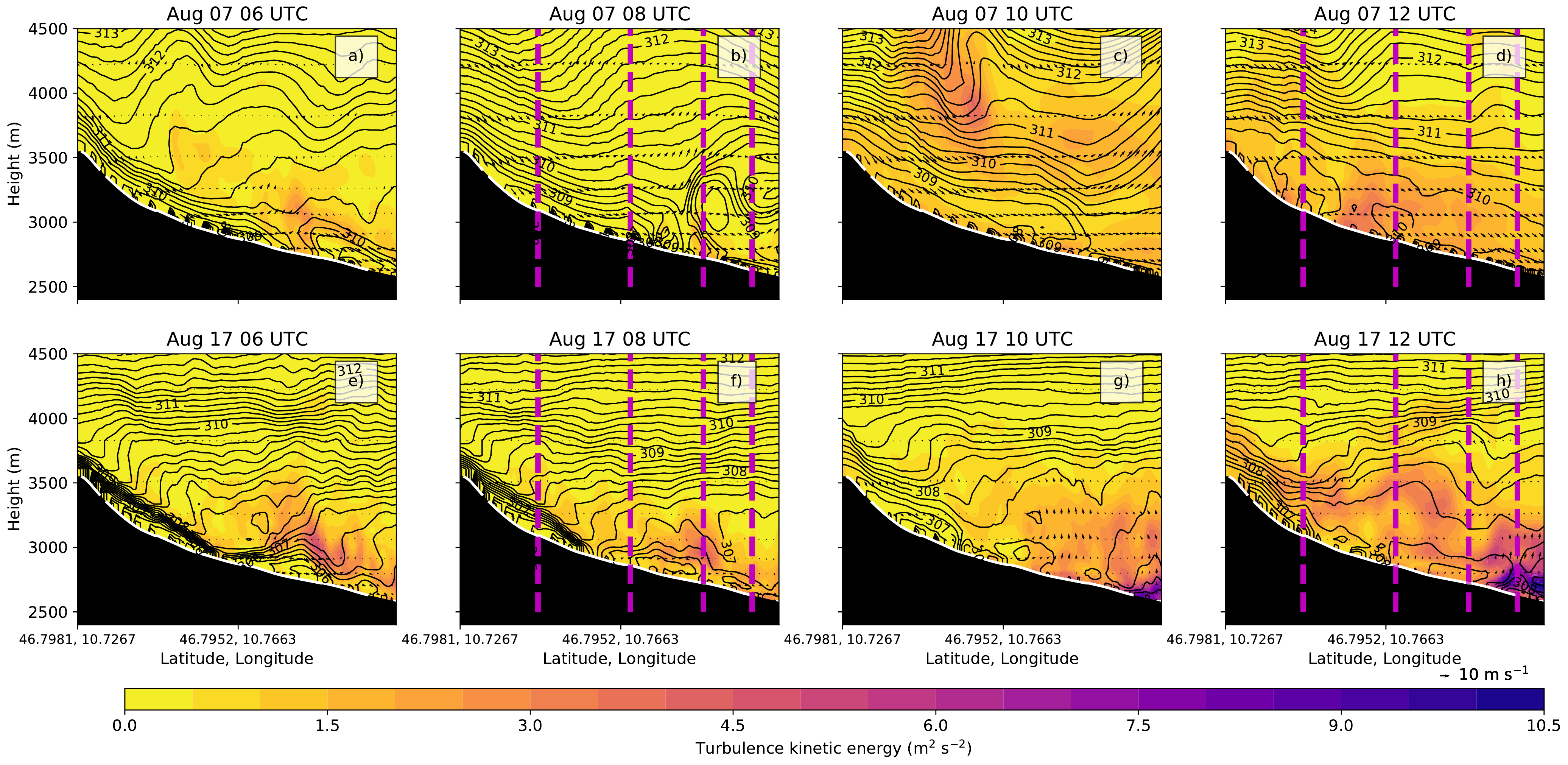}
 \caption{As Fig.~4, but along the black line in Fig.~1, and wind vectors indicate the along-valley wind speed. The vertical dashed lines show the segments of the cross-section and the points where vertical profiles are extracted for Fig.~6.}
\label{fig5}
\end{figure}

\begin{figure}[bt]
\centering
\includegraphics[width=0.5\textwidth,keepaspectratio]{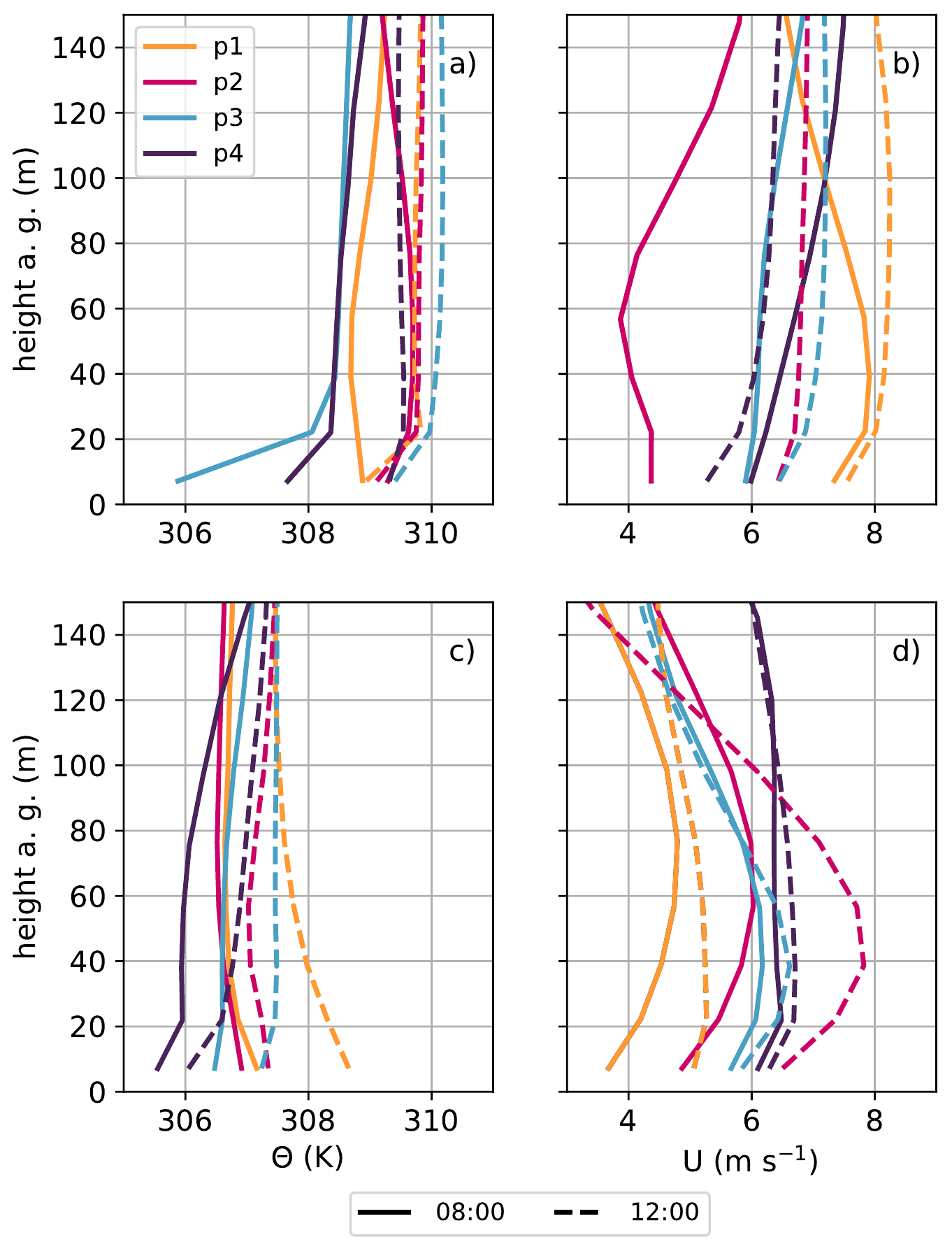}
\caption{Vertical profiles from model output of the SW day (a-c) and the NW day (d-f) at 08\,UTC (full lines) and 12\,UTC (dashed lines) of potential temperature (a,d), horizontal wind speed (b,e), and the Scorer parameter (c,f). The colors show the four different locations (dashed magenta lines in Fig.~5). The lines start at 7\,m above ground (the lowest model half-level).  }\label{fig6}
\end{figure}

\begin{figure}[bt]
\centering
\includegraphics[width=0.5\textwidth,keepaspectratio]{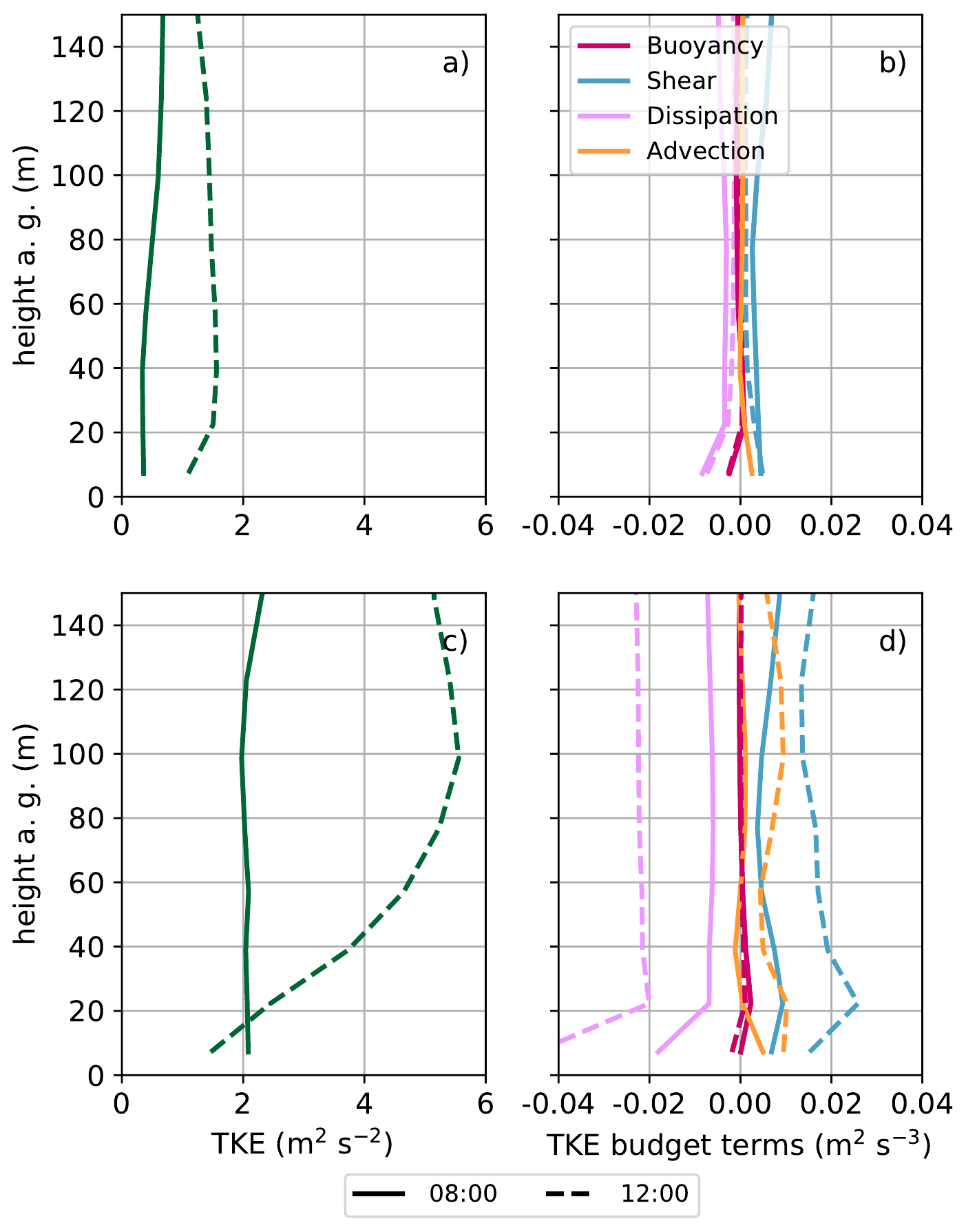}
\caption{Vertical profiles from model output of the SW day (a,b) and the NW day (c,d) at 08\,UTC (full lines) and 12\,UTC (dashed lines) of resolved turbulence kinetic energy (a,c), and the TKE budget terms (b,d).  The lines start at 7\,m above ground (the lowest model half-level).  }\label{fig7}
\end{figure}

Alpine glaciers are not isolated from their environment, because the surrounding complex terrain has a strong influence on stationarity, heterogeneity, and the general turbulence structure. We explore the related atmospheric processes with vertical cross-sections of potential temperature, the wind vectors, and turbulence kinetic energy (TKE) across the glacier valley (Fig.~\ref{fig4}). On the SW day at 06\,UTC, the the potential temperature field shows stable stratification directly over the glacier surface with large local differences (Fig.~\ref{fig4}a). The minimum in potential temperature coincides with the body of potentially colder air in Fig.~\ref{fig3}b,c. These local stable layers directly over the glacier surfaces remain present for the next hours (Fig.~\ref{fig4}b,c), but the stratification weakens subsequently, which is also visible in the spatial structure of potential temperature in Section~\ref{sec_pottemp}. TKE maxima are present above glacier surfaces, suggesting locally-generated turbulence, but also at upper levels, related to possible interaction of the larger-scale flow with the topography. On the NW day the atmosphere in the glacier valley is strongly stably stratified (Fig.~\ref{fig4}e,f). This stable stratification, together with the NW synoptic flow, favors the formation of a gravity wave in the vicinity of the ridge, north-west of HEF. The isentropes subsequently steepen, together with a visible elevated maximum of TKE around 200\,m over the glacier (Fig.~\ref{fig4}e,f). Around 12\,UTC, the gravity wave breaks (Fig.~\ref{fig4}h), and strong mixing and severe turbulence with TKE values higher than 10\,m$^2$\,s$^{-2}$ are present over the glacier tongue. The non-stationary, breaking gravity wave in the lee of the upstream steep topography (Fig.~\ref{fig4}h), and is the main reason for the continuous change in wind speed and direction in the timeseries of the NW day (Fig.~\ref{fig2}d,h). \newline
The along-glacier cross-section shows that gravity wave-like structures are also present on the SW day from the early morning at upper levels (Fig.~\ref{fig5}a). However, the atmosphere aloft is not as stable as on the NW day, and the slope along the glacier is less steep than the North-Western slope, one of the factors (asides mountain height and stratification) being less favorable for gravity wave formation. At 10\,UTC (Fig.~\ref{fig5}c), gravity wave breaking is visible at levels 1000\,m above the glacier surface, however, HEF is not affected by these upper-level gravity waves, and the local glacier boundary layer remains stably stratified with minima of potential temperature over the glacier surface (Fig.~\ref{fig5}a,b,c,d). On the NW day, the along-glacier cross-section reveals strong stability over the accumulation area of HEF above 3000\,m (Fig.~\ref{fig5}e,f), which is not present at the glacier tongue. The major reason for this weakening is likely the strong cross-glacier flow related to the gravity wave, eroding the local glacier boundary layer at the glacier tongue. When the gravity wave breaks, the strong mixing entirely dominates the glacier tongue, while a weak stable layer remains above the accumulation zone of HEF (Fig.~\ref{fig5}g). These two case studies show that background flow direction, atmospheric stability, and the upstream topography have a relevant impact on the formation and strength of gravity waves over the glacier valley. Furthermore, these factors determine if the glacier tongue is affected by gravity-wave breaking and the related severe turbulence.\newline
The cross-sections from the previous section gave an overview of the atmospheric processes contributing to the glacier boundary layer structure. In the following paragraphs, we have a more detailed look at the thermal stratification and wind speed with vertical profiles (Fig.~\ref{fig6}) from specific points over the glacier (highlighted in Fig.~\ref{fig5}). On the SW day at 08\,UTC (Fig.~\ref{fig6}a), the potential temperature profile reveals a  stable stratification at the lowest 20\,m at all chosen points, while the upper points on the glacier (p3 and p4) show a stronger stability than the glacier tongue points (p1 and p2). The stable stratification remains present until 12\,UTC at all points (dashed lines), while a neutral layer is present above the near-surface stable layer. The vertical profile of horizontal wind speed (Fig.~\ref{fig6}a) suggests mainly constant wind speeds with height, although a very weak local maximum is present at around 20\,m, at the same height as the stable layer in the potential temperature profile. The synoptic wind direction is the same as the down-glacier flow, therefore we can assume that the down-glacier wind is supported by the synoptic flow. The strongly stable stratification and the very weak jet maximum hint that the down-glacier flow might have partly katabatic origin. Judging from the vertical profiles, we can not decide definitely whether the forcing of the local layer above the glacier is katabatic in the model. The simulated local stable layer is not entirely comparable to observed katabatic flows during the HEFEX campaign with a wind speed maximum of 2\,m above ground \citep[][their Fig.~4]{Mott.etal.2020_Spatio-temporal}. The lowest model half-level is located at 7\,m, while the lowest 100\,m of the atmosphere in the model consist of seven model levels. Thus, it is likely that the model is not able to resolve shallow katabatic glacier winds with the vertical grid. \newline
On the NW day (Fig.~\ref{fig6}d-f), the vertical profiles show a different picture: According to the potential temperature profile at the two lower points p1 and p2 (08\,UTC, glacier tongue), the atmosphere above the glacier surface is neutrally stratified, while at the two points in the accumulation area of HEF (p3 and p4), the potential temperature profiles suggest a stable layer above the glacier ice as on the SW day. Atmospheric stability decreases with time, while at 12\,UTC, a slightly unstable layer is present above the glacier tongue. The vertical profiles show that a stable boundary layer over the glacier (profiles from p3 and p4 at 08\,UTC) coincides with a weak jet maximum. While the stable layer in the accumulation area of HEF remains persistent until 12\,UTC, the gravity wave and the induced cross-glacier flow erode the glacier boundary layer and strong turbulent mixing leads to a neutral stratification over the glacier tongue (p1 and p2). This suggests that even dynamically-induced atmospheric processes such as the gravity wave can be very localized, and that atmospheric conditions at the glacier tongue are not automatically representative for the entire glacier.\newline
The vertical structure of resolved turbulence kinetic energy (TKE) and its budget terms also give insights into the vertical ABL structure over the glacier tongue at hefex-3 (Fig.~\ref{fig7}). The TKE budget equation can be written after \cite{stull_turbulence_1988-1}:
 \begin{equation}\label{eq:tke_obs}
  \underbrace{\frac{ \partial \bar{e}}{ \partial t}}_\text{local tendency}+\underbrace{\overline{U_j}\frac{ \partial \overline{e}}{ \partial x_j}}_\text{advection}=\underbrace{{\delta_{i3}\frac{g}{\overline{\theta_v}}(\overline{u_i'\theta_v'})}}_{\substack{\text{buoyancy} \\ \text{production/consumption}}}-\underbrace{{\overline{u_i'u_j'}\frac{ \partial \overline{U_i}}{ \partial x_j}}}_{\substack{\text{shear}\\ \text{production}}}-\underbrace{\frac{ \partial(\overline{u_j'e})}{ \partial x_j}}_\text{turbulent transport }-
  \underbrace{\frac{1}{\overline{\rho}}\frac{ \partial (\overline{u_i'p'})}{ \partial x_i}}_\text{pressure correlation}-\underbrace{{\varepsilon}}_\text{dissipation}
 \end{equation} where capital letters with overbars denote mean quantities, while small letters with primes refer to turbulent fluctuations; $\bar{e}$ is TKE, $U$ is the mean wind speed, $g$ is the acceleration due to gravity, $\theta_v$ is virtual potential temperature, $\rho$ is air density, and $p$ is pressure. On the left-hand side (l.h.s.) are the local TKE tendency and advection with the mean flow, while on the right-hand side (r.h.s.) we find the thermally-driven buoyancy production/consumption term, the mechanical shear production term, the turbulent transport, the pressure correlation term, and the TKE dissipation rate $\varepsilon$. For our analysis, we consider the TKE advection, the vertical buoyancy production, the shear production, and the dissipation rate from the averaged model output. We have to note that in sloping terrain horizontal contributions to the buoyancy production might be non-negligible \citep{Oldroyd.etal_2016_Buoyant}, but we can not assess them with our present model set-up. Turbulent transport and the pressure correlation term mainly serve as a redistribution mechanism of TKE, but we do not consider them in our analysis. \newline 
 On both case study days, the TKE magnitude increases between 08\,UTC and 12\,UTC (Fig.~\ref{fig7}a,c). On the SW day, the vertical profile of TKE is almost constant with height at 08\,UTC, while at 12\,UTC, a local TKE maximum is present at around 20\,m, coincident with the stable layer (Fig.~\ref{fig6}a,b). The corresponding TKE budget terms suggest that the negative buoyancy term acts as a sink of TKE close to the ice surface, and that shear production is the major contributor to the TKE budget, counteracted by the dissipation, implying that the processes contributing the TKE are mostly local. \newline
 On the NW day, the vertical profile of TKE at 08\,UTC is also constant with height and the TKE budget structure is similar to the SW day. However, at 12\,UTC the TKE values increase dramatically up to 6\,m$^2$\,s$^{-2}$, also at upper levels. The TKE budget terms suggest again that shear is the dominant production term with values up to 0.02\,m$^2$\,s$^{-3}$, while the buoyancy term is closer to zero. However, the TKE advection is of almost equal magnitude as the shear production term at upper levels. During strong dynamically-induced situations with breaking gravity waves, the TKE advection is a non-negligible term of the TKE budget \citep[e. g.,][]{Vecenaj.etal_2010_Near,VecenajEtAl_2012_CoastFeaturesBora}. The NW day TKE structure is therefore not only influenced by local processes acting in the vertical, but also by larger-scale three-dimensional dynamics, visible in the generally high TKE values, but also in the TKE budget structure with TKE advection with the mean flow being a relevant term.
\section{Boundary Layer and Surface Fluxes}\label{sec4}
\subsection{Sensible Heat Flux Structure}
\begin{figure}[bt]
\centering
\includegraphics[width=\textwidth, keepaspectratio]{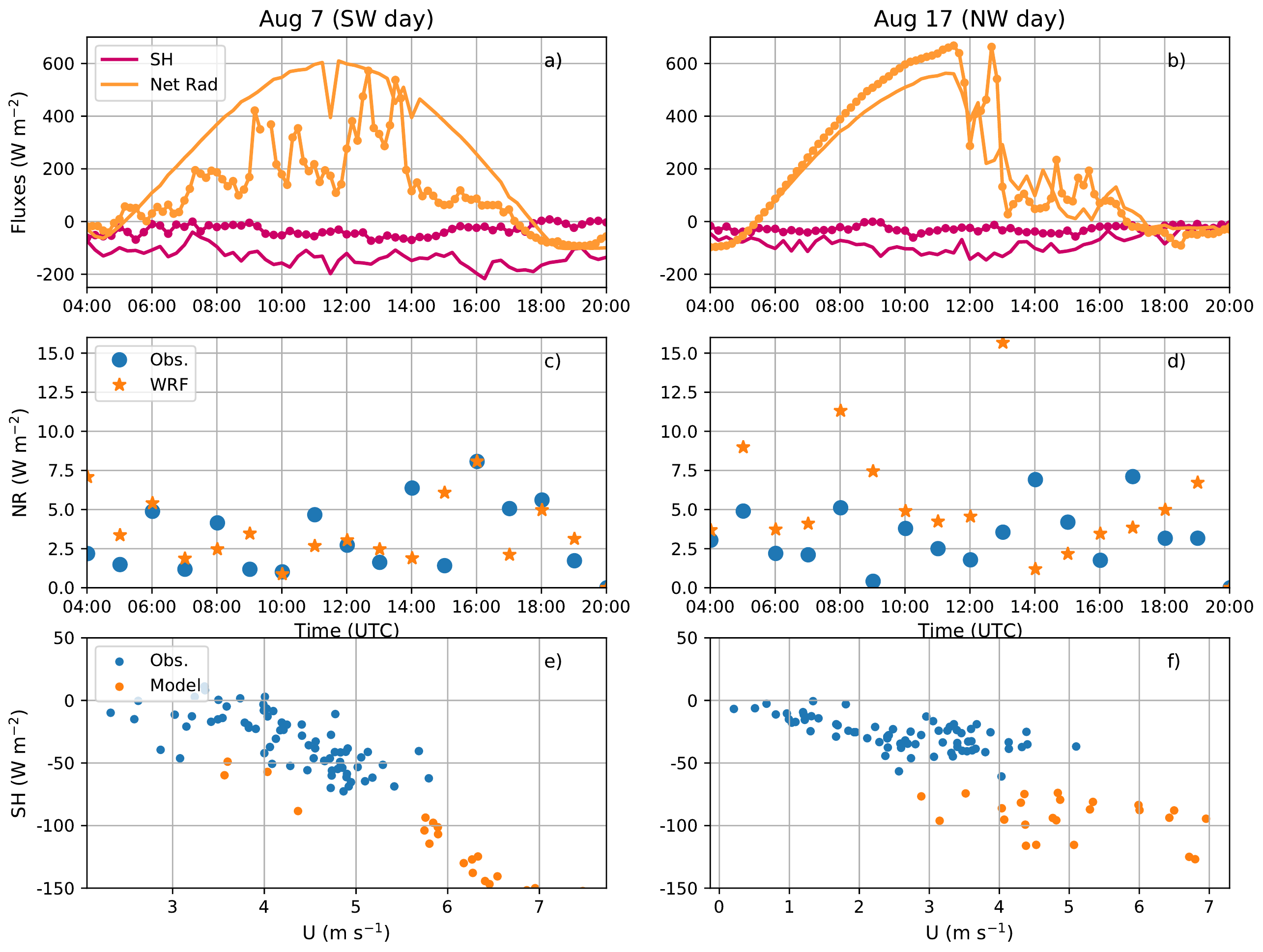}
\caption{(a,b) Time series of the net radiation from StHE (orange lines) and the surface sensible heat flux of hefex-3 (pink lines) for both case study days, where straight lines show model output and lines with dots show observations. (c,d) $NR$ of the sensible heat flux time series of one-minute observations (blue points) and one-minute model output (orange stars). (e,f) Scatter plots of horizontal wind speed and sensible heat flux from observations (blue) and model output (orange) between 06\,UTC and 12\,UTC.}\label{fig8}
\end{figure}

\begin{figure}[bt]
\centering
\includegraphics[width=\textwidth,keepaspectratio]{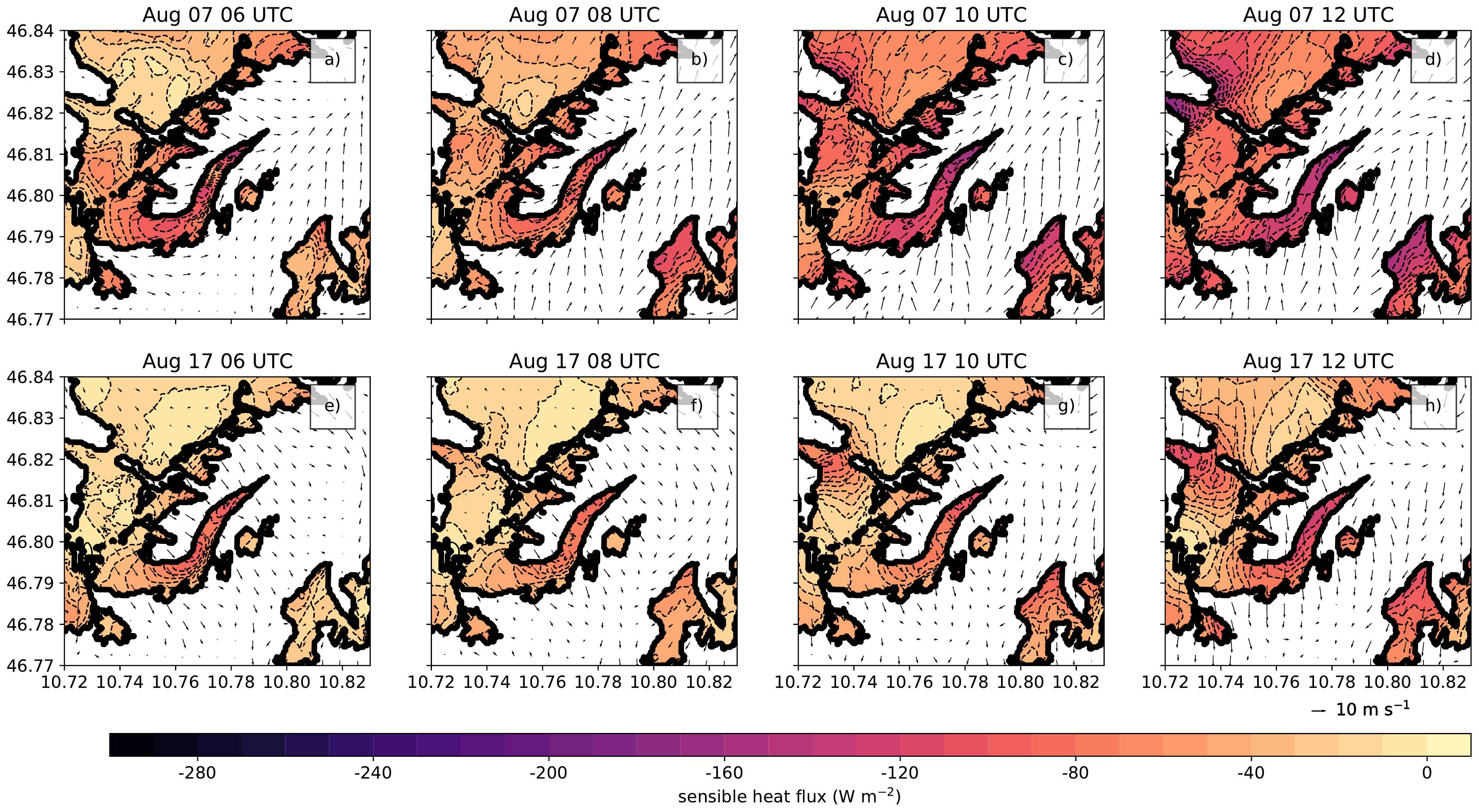}
\caption{Averaged model output of the surface sensible heat flux (colors and dashed lines) and horizontal wind vectors from the lowest model level over glaciated surfaces. Sensible heat fluxes over ice-free surfaces are not shown. The left column shows data from the SW day, while the right column shows data from the NW day, at four different times.}\label{fig9}
\end{figure}

The previously described meteorological phenomena are spatially heterogeneous, and especially the wind speed and direction might have an impact on sensible heat flux variability over the glacier. However, according to the energy balance, sensible heat flux is mostly determined by the net radiation. Timeseries of net radiation from the SW day suggest some cloudiness, while the model simulates an almost cloud-free day (Fig~\ref{fig8}a). This also manifests itself in the sensible heat flux observations, which are generally smaller than the simulated sensible heat flux. The sign of the sensible heat flux is negative both in the observations and model output, suggesting that the ice surface is represented correctly in the model. The net radiation on the NW day shows a better agreement between the model and observations. The weather conditions are cloud-free before noon, while clouds establish in the afternoon. The sensible heat fluxes are smaller than on the SW case study day both in the observations and the model, but the sign is also negative over the glacier surface.\newline
The animation from one-minute model output [see additional material] shows a very high temporal variability of the sensible heat flux. For the SW day, with increasing daytime turbulence, plumes of local sensible heat flux maxima travel down the glacier tongue with the mean wind. These plumes prevail for the rest of the day, suggesting a high spatio-temporal variability. As for the NW day, the breaking gravity wave is the largest driver of high temporal variability in sensible heat flux. This raises the question whether this non-stationarity is also present in the observations, and if it has a diurnal cycle. For this purpose, we divide the one-minute sensible heat flux time series (from both model output and observations) into 15-mintue intervals and calculate the non-stationarity ratio after \cite{Mahrt_1998_FluxSamplingErrors}:
\begin{equation}
 NR=\frac{\sigma_{k,btw}}{RE_k},
\end{equation}
where $\sigma_{k,btw}$ is the variance of each of the sub-intervals and $RE_k$ is the standard error of the 15-minute averaged flux results. When the $NR$ is approximately unity, the time series can be assumed to be stationary, because the variances of the 15-minute averages do not exceed their standard error. For values higher than 2, non-stationarity can be assumed. We are aware that the non-stationarity depends on the choice of the averaging period, but similar averaging choices were made for the $NR$ of sensible heat fluxes under gravity-wave influence during the T-REX experiment \citep{Vecenaj.DeWekker_2015_Determination}. \newline 
Figure~\ref{fig8}c,d shows the $NR$ of the sensible heat flux from the two case study days. One-minute instantaneous values for both days from model and observations show mainly values around or above 2, suggesting non-stationary behavior. While the $NR$ gradually increases with increasing daytime turbulence on the SW day, the $NR$ on the NW day is in general higher without a diurnal cycle: $NR$ maxima are already present at 08\,UTC, possibly related to the strong gravity-wave activity which continues until 12\,UTC (cf. Fig.~\ref{fig4}f). Furthermore, since the $NR$ is higher in the model than in the observations, the model seems to simulate a stronger gravity wave influence. The higher the non-stationarity, the more mesoscale contributions influence the surface sensible heat flux. The high non-stationarity in the time series might partly explain the large differences between observed and modelled sensible heat flux.  \newline  
Asides from temperature, the horizontal wind speed also influences the strength of sensible heat fluxes (Fig.~\ref{fig9}e,f): The higher the horizontal wind speed, the stronger the sensible heat flux; it has to be mentioned at this point that the model tends to overestimate the horizontal wind speed, therefore also producing higher sensible heat fluxes than observed. Although there are differences in absolute magnitude between the model and the observations, the linear relationship between the wind speed and the sensible heat flux is similar, suggesting that the model is able to reproduce the relevant physical processes. To conclude, the magnitude of the sensible heat flux is strongly governed by the horizontal wind speed. \newline 
The observations only show the sensible heat flux at a point, but the model output suggests that the averaged sensible heat flux over the glacier surface is spatially highly heterogeneous on both case study days. On the SW day at 06\,UTC (Fig.~\ref{fig9}a), the spatial sensible heat flux structure is connected to the horizontal wind speed and direction, while the weak cross-wind with potentially warmer air from the slopes is also visible with the higher sensible heat flux over the glacier tongue. This changes at 08\,UTC (Fig.~\ref{fig9}b), visible with weaker sensible heat fluxes over the glacier tongue and streaks in the structure, connected to the body of potentially colder air  and the down-glacier wind (Fig.~\ref{fig3}b). At 10\,UTC, the streaks are still visible, but the sensible heat flux is now stronger; this is connected to the weakening stratification over the glacier tongue. At noon (Fig.~\ref{fig9}d), together with the increasing horizontal wind speed, the sensible heat flux over the glacier tongue reaches its maximum. On the NW day, the sensible heat fluxes are generally smaller than on the SW day, but they show an atypical pattern: At 06\,UTC, a local sensible heat flux maximum is already present at the glacier tongue downwind from the south-facing slope (Fig.~\ref{fig9}e). This structure is visible for the next six hours (Figs.~\ref{fig9}f-h). When at 12\,UTC the gravity wave breaks and severe turbulence is present over the glacier tongue, the sensible heat flux is strongest (Fig.~\ref{fig9}h and Fig.~\ref{fig9}b). While on the SW day the persistent down-glacier flow leads to a specific spatial structure (streaks) over the glacier tongue, the strong mesoscale influence with the gravity wave also leads to localized high sensible heat flux values over the glacier tongue only.
\subsection{Temperature Advection Patterns and Heat Budget}
\begin{figure}[bt]
\centering
\includegraphics[width=\textwidth,keepaspectratio]{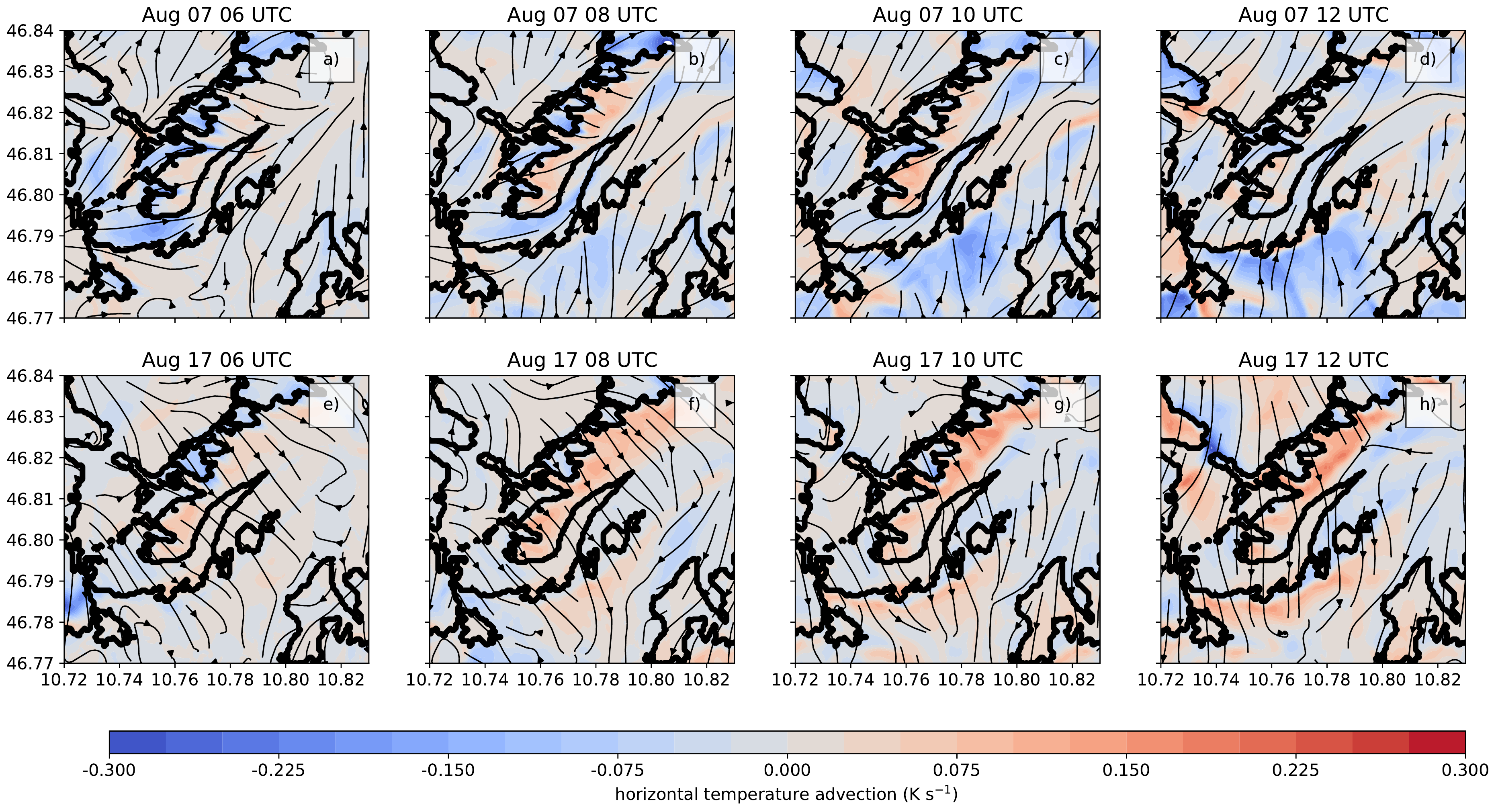}
\caption{Averaged model output of the averaged horizontal temperature advection (colors) and streamlines derived from the wind field from the lowest model level. The left column shows data from the SW day, while the right column shows data from the NW day, at four different times.}\label{fig10}
\end{figure}

\begin{figure}[bt]
\centering
\includegraphics[width=0.5\textwidth,keepaspectratio]{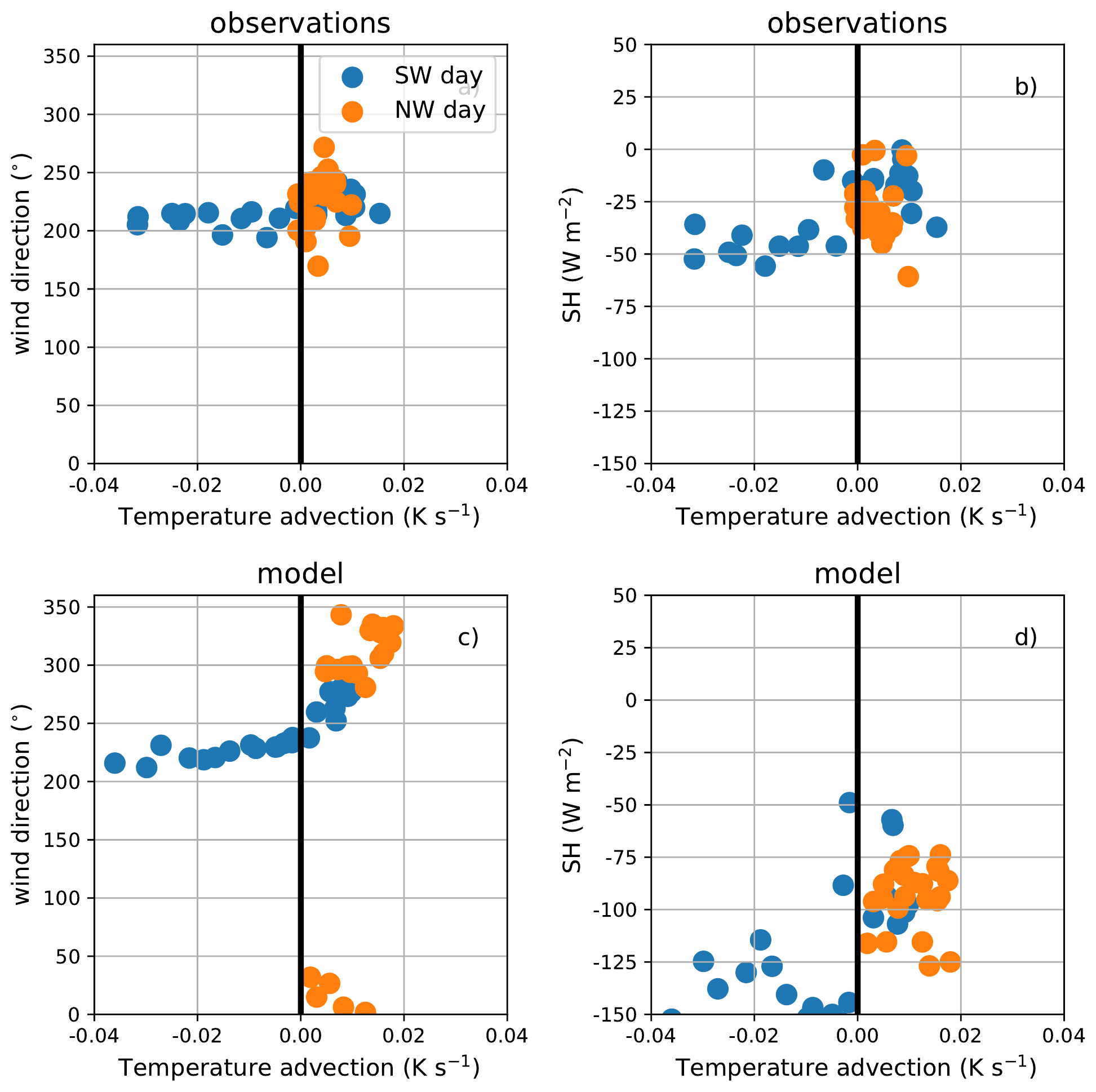}
\caption{Scatter plots of the SW day (blue) and the NW day (orange) from observations (upper row) and model output (lower row) of location hefex-3 for horizontal temperature advection and wind direction (a,c), and horizontal temperature advection and sensible heat flux (b,d).}\label{fig11}
\end{figure}

\begin{figure}[bt]
\centering
\includegraphics[width=0.5\textwidth,keepaspectratio]{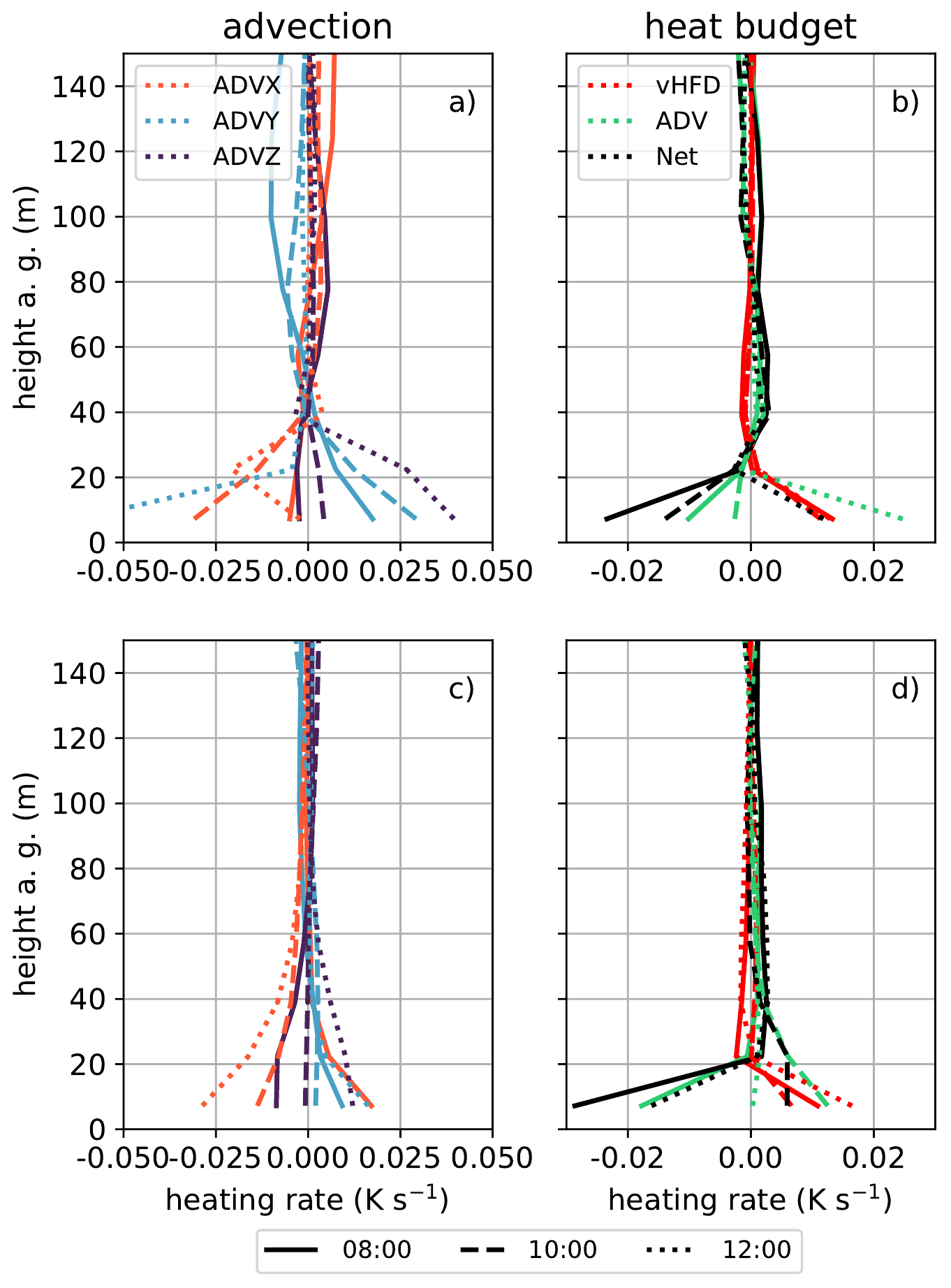}
\caption{Vertical profiles from model output of the SW day (a,b) and the NW day (c,d) at 08\,UTC (full lines), 10\,UTC (dashed lines), and 12\,UTC (dotted lines) from hefex-3 of the advection terms (a,c), and the heat budget (b,d).  The lines start at 7\,m above ground (the lowest model half-level).}\label{fig12}
\end{figure}

The heterogeneous spatial structure of the wind field and potential temperature determine the spatial patterns of the horizontal temperature advection (Fig.~\ref{fig10}). On the SW day, the body of potentially colder air (Fig.~\ref{fig3}) coincides with horizontal cold air advection over the glacier. At 06\,UTC, when the colder air is pushed towards the North-facing slope, the glacier tongue is under the influence of weak warm-air advection. This changes at 08\,UTC, when the down-glacier wind increases and the glacier tongue is under persistent cold-air advection. The streamlines suggest that the ``source'' of the cold air is mainly located at the glacier accumulation area. Until 12\,UTC, the thermal stratification decreases (Fig.~\ref{fig6}a), but the glacier tongue remains under the influence of cold-air advection. This suggests that if the wind direction is down-glacier, the glacier remains in the presence of cold-air advection. The NW day shows a different temperature advection pattern: The cross-glacier circulation leads to advection of warmer air over the glacier tongue. The absolute values of the warm-air advection over the glacier are, however, much smaller than the absolute values of the cold-air advection of the SW day. The wind direction determines the type of advection over the glacier tongue: The upstream source region is the heated slope, and, supported by the gravity wave, the glacier tongue is under the influence of, albeit weak, warm-air advection. On the other hand, the glacier accumulation area, where local stable stratification is present (Fig.~\ref{fig6}d) remains under very weak cold-air advection.  \newline
To explore the relation between the horizontal temperature advection and the wind speed, we compare the model output with observations from the EC stations. To examine the impact of cross-glacier flows on horizontal temperature advection, we calculate the observed horizontal temperature advection between stations hefex-3 and hefex-2:
\begin{equation}
 THADV=U\frac{\Delta T}{\Delta x},
\end{equation}
Where $\Delta x=179$\,m is the distance between the stations, $\Delta T$ is the temperature difference, and $U$ is the average horizontal wind speed. The observations (Fig.~\ref{fig11}a) suggest that the glacier tongue is under the influence of horizontal cold-air advection. The wind direction is mostly down-glacier at 200$^{\circ}$, but when deviates towards 250$^{\circ}$, the glacier tongue is under weak warm-air advection, and the same patterns are visible in the model output  (Fig.~\ref{fig11}c). On the NW day, both observations and the model suggest that the glacier tongue is under continuous warm-air advection. This supports the idea that if a down-glacier wind is present (corresponding to a wind direction 200$^{\circ}$), horizontal cold-air advection dominates, but if the wind direction shifts and the source area therefore changes, warm-air advection dominates on the glacier tongue. Asides from the wind direction shift in the model (discussed in Sec.~\ref{sec3}), the magnitude of the horizontal temperature advection is similar in the observations and the model. In general, the cold-air advection connected to the down-glacier flow on the SW day is stronger than the warm-air advection under intermittent conditions on the NW day, also observed by \cite{Mott.etal.2020_Spatio-temporal}. The relation between the horizontal wind speed and the resulting temperature advection are shown in Fig.~\ref{fig11}b,d. The pattern of the scatter is similar in the observations and the model, while the overestimated wind speeds in the model again lead to a shifted patterns towards higher wind speeds. However, the strength of the advection seems to be related to the wind speed strength, e.g. strong cold-air advection the SW day is related to higher wind speeds. On the other hand, the type of advection (cold-air or warm-air advection) is then rather related to the horizontal temperature gradient. \newline
Horizontal temperature advection is only one term of the temperature tendency equation in the boundary layer \citep{Wyngaard_2010_Turbulence,Haid.2020_Foehncoldpoolinteractions}:
\begin{equation}
 \frac{\partial \theta}{\partial t}=\underbrace{-U\frac{\partial \theta}{\partial x}-V\frac{\partial \theta}{\partial y}-W\frac{\partial \theta}{\partial z}}_{Adv}-\underbrace{\frac{\partial \overline{w' \theta'}}{\partial z}}_{vHFD}
\end{equation}
Consisting of temperature advection with the mean wind in all three directions (ADVX, ADVY, ADVZ) and the vertical heat flux divergence (vHFD), while the radiative flux divergence is neglected. Vertical profiles of the three temperature advection terms (Fig.~\ref{fig12}a) from the SW day at 08\,UTC and 10\,UTC suggest that cold-air advection from the x- and z- direction dominates over warm-air advection from the y-direction. Together with the decreasing vHFD over hefex-3 this results in a net cooling of the atmosphere over the glacier tongue (Fig.~\ref{fig12}b). However, at 12\,UTC, the net positive advection contribution in the heat budget (Fig.~\ref{fig12}b), leads to a net warming of the atmosphere over the glacier. This heat budget analysis shows that a down-glacier wind with lateral cold-air advection is not a guarantee for general cooling of the atmosphere over the glacier. On the NW day, cold-air advection dominates the glacier tongue at 08\,UTC (Fig.~\ref{fig12}c), with the dominant advection from the x-direction. As on the SW day, this leads to a net cooling at 08\,UTC (Fig.~\ref{fig12}d). Two hours later (10\,UTC), the advection part of the heat budget turns positive, leading to an overall warming tendency over the glacier tongue. At 12\,UTC, the situation changes: while the overall advection turns positive, the overall heat budget remains negative. The breaking gravity wave mixes potentially colder air from aloft (also supported by the potentially colder air over the large glacier upstream of HEF) into the glacier boundary layer. This heat budget analysis shows that it is not possible to judge only from the lateral advection components on the atmospheric warming or cooling above the glacier, because both the vertical temperature advection and the vHFD are terms of equal magnitude. Therefore, our numerical modelling approach gives useful insights to the individual components of the heat budget over the glacier.
\section{Discussion}\label{sec5}
The simulation of boundary-layer processes in complex terrain is still a major challenge for NWP models \citep{Edwards.etal_2020_Representation}, due to insufficient terrain representation \citep{Wagner.etal_2014_Impact}, too coarse land-surface datasets \citep{GolzioEtAl_2021_LandUseImprovements} or turbulence parameterizations developed based on assumptions for horizontally homogeneous and flat terrain \citep{Goger.etal_2018_Impact, Goger.etal_2019_New}. These problems can partly be avoided by increasing horizontal and vertical resolution towards horizontal grid spacings where the largest eddies can be expected to be already resolved \citep{Chow.etal_2019_Crossing}. Our LES set-up at $\Delta x=48$\,m has a realistic representation of the topography surrounding the glacier (within the scope of numerical stability), and furthermore, the horizontal mesh size also allows a realistic representation of the land-use and the ice surface (Fig.~\ref{fig1}). \newline 
Other studies comparing LES with WRF with high-resolution observations suggest good model performance, so that the model can be used as a tool for better process understanding \citep{Munoz-Esparza.etal_2017_Coupled,Babic.DeWekker_2019_Characteristics,Hald.etal_2019_Large,Conolly.etal_2020_Nested}. However, in our complex terrain setting the model still overestimates the horizontal wind speeds on both days. One reason for this are the differences in measuring heights of the instruments (2\,m) versus the lowest model level at 7\,m. Another reason for the overestimated wind speeds can be attributed to the still insufficient topography representation, although \cite{Wagner.etal_2014_Impact} point out that relevant boundary-layer processes are resolved when the valley is represtented by at least 10 grid points, which is the case for the innermost domain in our LES. In general, other WRF-LES studies over truly complex terrain show a similar bias of overestimated wind speeds \citep{Gerber.etal_2018_Spatial,Liu.etal_2020_Simulation,Umek.etal_2021_Large}.
However, model topography still differs from reality due to smoothing, because the development and formation of the thermally-induced circulation, katabatic winds, boundary-layer separation, and gravity waves also depend on the topography steepness and/or shape \citep{Wagner.etal_2015_impact,Smith.Skyllingstad_2005_Numerical,PrestelWirth_2016_WhatFlowConditions,Doyle.etal_2011_Intercomparison}. Furthermore, the restriction of model topography not having slope angles over $40^{\circ}$ also reduces complexity of the ABL flow (e.g. no bluff-body boundary-layer separation in the lee of very steep mountaintops). This well-known problem of vertical grids over complex terrain might be solved with the implementation of the immersed-boundary method in the near future \citep{Lundquist.etal_2010_Immersed}.  \newline 
Our current set-up is not able to resolve all relevant length scales equally well. For example, on the NW day, the synoptic wind direction is well-simulated at the ridge station IHE (Fig.~\ref{fig2}f), and the model simulates a strong, breaking gravity wave over the glacier tongue. The gravity wave easily erodes the glacier ABL in the simulation, while the observations suggest a weak, but strongly disturbed katabatic wind layer. This hints that the model simulates a stronger gravity wave influence than found in reality, also visible in the high TKE values (more than 10\,m$^2$\,s$^{-2}$ in Fig.~\ref{fig4}h) and the higher non-stationarity of the sensible heat flux than in the observations (Fig.~\ref{fig11}d). The HEFEX stations did not record such high TKE values over the glacier tongue on the NW day, but longer TKE time series from station IHE reveal that comparably high TKE values (around 10\,m$^2$\,s$^{-2}$) are possible in connection with NW synoptic flows. Furthermore, aircraft observations by \cite{JiangDoyle_2004_GravityWaveBreaking} over the \"{O}tztal Alps suggest that our simulated TKE values (around 10\,m$^2$\,s$^{-2}$) are realistic. The predictability of mountain waves decreases with increasing non-linearity and terrain steepness \citep{Doyle.etal_2011_Intercomparison}, and strong gravity waves are often a challenge for the simulation of dynamically-driven flows over mountainous terrain \citep{Gohm.etal_2004_South,Umek.etal_2021_Large}. \newline 
One of the largest questions when setting up the model was whether WRF-LES is able to simulate small-scale glacier boundary-layer processes. \cite{Cuxart_2015_When} states that a horizontal grid spacing of $\Delta x=5$\,m might be sufficient to simulate processes in the stable mountain boundary layer. Although we use a coarser grid spacing, our simulations suggest that the representation of the glacier boundary layer is partly possible: On both days, the model simulates a stably stratified layer over the ice surfaces, which remains persistent during the daytime (SW day) or is eroded by the gravity wave at the glacier tongue (NW day). The outer domain (d03, $\Delta x$=240\,m) is not able to simulate the stable layer over the ice surfaces (not shown), although the vertical grid spacing is the same. This implies that the horizontal mesh size has an essential influence on the correct simulation of the the spatial variability of ABL processes over the glacier. \newline 
The different scales in the ABL over the glacier are not isolated from each other and scale interactions are common in both case study days. Small-scale processes over the glacier, such as local stable layers, are still underrepresented in the model, and therefore less resilient towards meso-scale influence, although they might be more persistent in reality. For example, a down-glacier katabatic wind during summer might have a jet maximum of only 2\,m above ground \citep[as in][]{Mott.etal.2020_Spatio-temporal}, which cannot be resolved by the model's vertical grid (lowest model half-level at 7\,m). However, especially at the SW day, the down-glacier wind is supported by the synoptic flow direction, resulting in weak jet maxima at around 20\,m. This is partly related to the model's vertical grid, but on the other hand \cite{Smeets.etal_1998_Turbulence} also observed deeper down-glacier winds  when the katabatic glacier wind was aligned with the synoptic flow, while low-frequency disturbances in the turbulence spectra were present. The influence of the synoptic flow direction on the turbulence structure was also observed over the Saint-Sorlin Glacier in the French Alps \citep{Litt.etal_2017_Surface}. Our results from the SW day agree with the aforementioned studies and show that the synoptic flow direction governs the spatial structure of the glacier boundary layer. Due to the glacier's high altitude, synoptic flows interact with the topography and plunge into the comparably shallow glacier valley, eroding the glacier boundary layer. This raises the question under which conditions a glacier is able to maintain its own microclimate. Purely thermally-driven situations (i.e., with an undisturbed glacier wind) as described in \cite{Smeets.etal_2000_Turbulence}, \cite{Nicholson.Stiperski.2020_Comparisonturbulentstructures}, and \cite{Mott.etal.2020_Spatio-temporal} are still possible under fair-weather conditions with weak background wind, but the synoptic wind direction and the local topography are major factors for the spatial sensible heat flux structure on the glacier and should not be disregarded.  \newline 
Asides from the aforementioned shortcomings, the model provides valuable information about the physical processes contributing to the glacier ABL structure. The simulation results agree well with the observational analysis of \cite{Mott.etal.2020_Spatio-temporal} for various quantities: The wind speeds over the glacier are higher when a down-glacier wind is present than during disturbed situations, the sensible heat flux is linearly dependent on the wind speed, and the connection between the wind direction and the horizontal temperature advection are present in the simulations. The large benefit of the simulations is, however, the information of spatial fields (both horizontal and vertical) of bpundary-layer variables such as the sensible heat flux, TKE, and temperature advection, which can not be observed even by an array of EC towers. The spatial field of the sensible heat flux especially shows that the elevation dependence of the sensible heat flux \citep{Greuell.etal_1997_Elevational} is only partially true for our simulation results: Non-linear processes such as gravity waves and the horizontal wind field over the glacier strongly contribute to non-stationarity and spatial heterogeneity of the sensible heat flux over the glacier. This raises questions whether linear relationships, single point measurements, and flux-profile methods  basing on Monin-Obukhov similarity theory are representative for surface energy balance modelling of mountain glaciers \citep{Denby.Greuell_2000_use, Sauter.Galos_2016_Effects}.

\section{Summary and Conclusions}\label{sec6}
We conducted large-eddy simulations with the WRF model at $\Delta x=48$\,m over the Hintereisferner glacier located in the \"{O}tztal Alps, Austria for two case studies with different synoptic forcings: On August 7, 2018 South-westerly flow was present, while on August 17, 2018 North-westerly flow prevailed. The model results are validated with EC towers on the glacier surface (HEFEX experiment) and permanent stations on surrounding ridges and/or slopes. Dependent on the direction of the synoptic flow, different wind patterns are present over the glacier affecting the spatial structure of the glacier boundary layer and the sensible heat flux. This leads us to the following conclusions:
\begin{enumerate}
 \item On the SW day, the model is able to simulate the temperature and wind patterns over the glacier in a satisfactory way. While the horizontal wind speed is generally overestimated by the model, the wind direction is simulated well for the SW day. This allows a further process-oriented analysis of the model output. On the NW day, the model simulates a strong cross-glacier wind, while the observed wind direction suggests a heavily disturbed down-glacier wind. However, the down-glacier wind from the observations is so shallow that the model can not resolve it with its current horizontal and vertical grid spacing.
 \item On the SW day, the simulated potential temperature structure reveals that a prevailing body of potentially colder air is present over the ice surfaces, accompanied by persistent down-glacier winds. On the NW day, strong cross-glacier flow leads to a more homogeneous potential temperature structure without horizontal temperature gradients between the glacier and surroundings. 
 \item Cross-sections along and across the glacier from the SW day reveal that small-scale processes such as differential heating are present and sources of TKE are likely local. Furthermore, the synoptic wind direction supports the down-glacier flow. On the other hand, on the NW day, a strong gravity wave forms on the upstream steep slope, resulting in severe shear-driven turbulence and an erosion of the boundary layer on the glacier tongue, while the accumulation area of the glacier is less affected.
 \item Vertical profiles of potential temperature and wind speed reveal that a stably-stratified layer is present over the glacier surface on the SW day. TKE values are generally low and weak shear is the only TKE production source. On the NW day, the layer above the glacier tongue is neutrally stratified without a distinguishable local layer, and the TKE values are high especially also at upper levels. This is connected to strong shear production, but also TKE advection acts as an important contributor to elevated TKE maxima.
 \item The model is generally able to simulate the net radiation and sensible heat flux structure over the glacier tongue. Both model and observations suggest high non-stationarity of sensible heat fluxes during the days. On the SW day non-stationarity is mostly related to a gradual increase in daytime turbulence, while on the NW day, the high non-stationarity of the sensible heat flux is related to the influence of the breaking gravity wave.
 \item On the SW day, together with the down-glacier wind, mainly horizontal cold-air advection is present over the glacier tongue, while on the NW day the cross-glacier flow induces mainly warm-air advection. The source region (i.e., either glacier accumulation zone or the surrounding slopes) influences whether cold-air or warm-air advection is present over the glacier tongue, while the horizontal wind speed mostly determines the advection strength. However, an overall analysis of the heat budget suggests that the lateral advection terms have a small influence on whether overall cooling or heating of the near-glacier atmosphere is present, because also the vertical temperature advection and vertical heat flux divergence play a relevant role.
 \item The synoptic flow, the upstream topography, and atmospheric stratification have a major influence on whether a local glacier boundary layer is able to form (SW day) or if it is completely eroded (NW day), while even dynamically-induced processes can be highly heterogeneous in space due to the underlying complex terrain.
 \item The horizontal and vertical resolution of our current model set-up is not yet high enough to resolve very shallow katabatic down-glacier winds. However, the model is successful in simulation of a daytime stable boundary layer over the glacier surfaces and the mesoscale processes which contribute to the glacier wind's spatial structure or possible erosion.
 
\end{enumerate}
 The Hintereisferner-LES set-up is a valuable tool to investigate boundary-layer processes over the glacier. The synoptic wind direction strongly governs the local glacier ABL formation and therefore it might be able to predict local conditions on the glacier when only the large-scale wind direction is known. Based on the findings from the simulations, an updated wind climatology over the glacier might give more insight on when a katabatic glacier wind is present or when synoptic winds dominate. Our simulations showed that the WRF-LES HEF set-up can be used for other case studies, and especially also for other seasons (i.e., winter) to deliver high-resolution wind fields for the analysis of wind-driven snow redistribution patterns \citep{VoordendagEtAl_2021_AUTOMATEDPERMANENTLONG} or for studies of WRF simulations coupled with energy balance models \citep{Sauter.eta_2020_COSIPY}. 

\section*{acknowledgements}
 This work is part of the project "Measuring and modeling snow-cover dynamics at high resolution for improving distributed mass balance research on mountain glaciers", a joint project fully funded by the Austrian Science Foundation (FWF; project number I 3841-N32) and the Deutsche Forschungsgemeinschaft (DFG; project number SA 2339/7-1). IS acknowledges FWF project T781-N32. The computational results presented have been achieved using the Vienna Scientific Cluster (VSC) under project number 71434.
\section*{conflict of interest}
The authors declare no conflict of interest.
\bibliography{references.bib}
\bibliographystyle{vancouver-authoryear} 
\newpage
\graphicalabstract{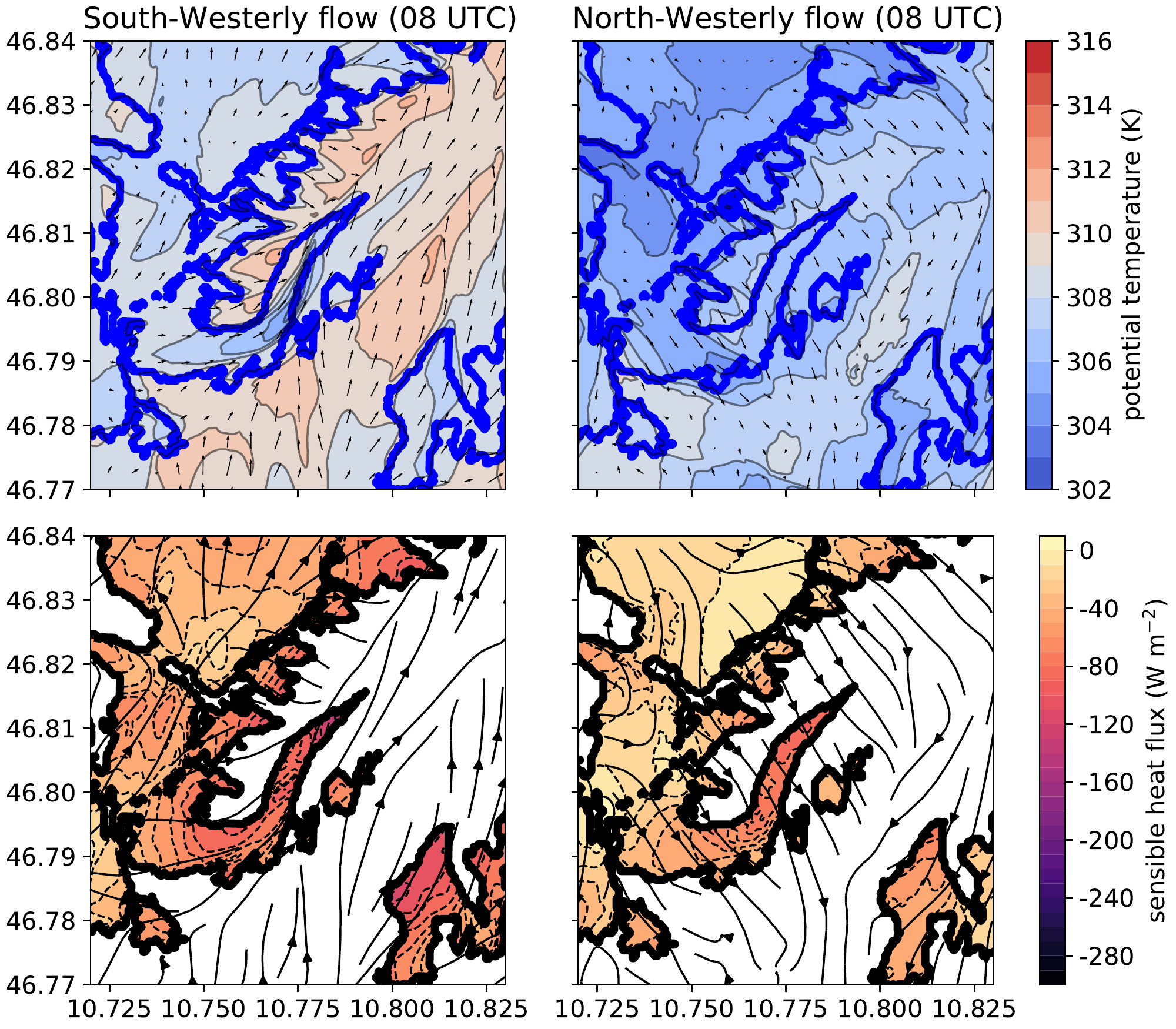}{}
\newpage




 \end{document}